\documentclass[acmlarge, screen]{acmart}

\usepackage{multirow}
\usepackage{makecell}
\usepackage{pifont}
\usepackage{color}
\usepackage{soul}
\usepackage{subfigure}

\AtBeginDocument{%
  \providecommand\BibTeX{{%
    \normalfont B\kern-0.5em{\scshape i\kern-0.25em b}\kern-0.8em\TeX}}}

\setcopyright{acmlicensed}
\acmJournal{IMWUT}
\acmYear{2023} \acmVolume{7} \acmNumber{3} \acmArticle{142} \acmMonth{9} \acmPrice{15.00}\acmDOI{10.1145/3610873}

\acmJournal{IMWUT}

\begin{document}

\title{Radio2Text: Streaming Speech Recognition Using mmWave Radio Signals}

\author{Running Zhao}
\authornote{Co-first author.}
\email{rnzhao@connect.hku.hk}
\orcid{0000-0003-2496-3429}
\affiliation{%
  \institution{The University of Hong Kong}
  \city{Hong Kong SAR}
  \country{China}
}

\author{Jiangtao Yu}
\authornotemark[1]
\orcid{0009-0004-3964-5874}
\email{jiangtyu2001@gmail.com}
\affiliation{
  \institution{Shanghai Qi Zhi Institute}
  \city{Shanghai}
  \country{China}}
\affiliation{
  \institution{IIIS, Tsinghua University}
  \city{Beijing}
  \country{China}}

\author{Hang Zhao}
\authornote{Corresponding author.}
\orcid{0000-0003-1928-7841}
\email{hangzhao@mail.tsinghua.edu.cn}
\affiliation{
  \institution{IIIS, Tsinghua University}
  \city{Beijing}
  \country{China}}
\affiliation{
  \institution{Shanghai Qi Zhi Institute}
  \city{Shanghai}
  \country{China}}

\author{Edith C.H. Ngai}
\authornotemark[2]
\orcid{0000-0002-3454-8731}
\email{chngai@eee.hku.hk}
\affiliation{
  \institution{The University of Hong Kong}
  \city{Hong Kong SAR}
  \country{China}
}

\renewcommand{\shortauthors}{Zhao and Yu, et al.}

\begin{abstract}
    Millimeter wave (mmWave) based speech recognition provides more possibility for audio-related applications, such as conference speech transcription and eavesdropping. However, considering the practicality in real scenarios, latency and recognizable vocabulary size are two critical factors that cannot be overlooked. In this paper, we propose \textit{Radio2Text}, the first mmWave-based system for streaming automatic speech recognition (ASR) with a vocabulary size exceeding 13,000 words. \textit{Radio2Text} is based on a tailored streaming Transformer that is capable of effectively learning representations of speech-related features, paving the way for streaming ASR with a large vocabulary. 
    To alleviate the deficiency of streaming networks unable to access entire future inputs, we propose the \textit{Guidance Initialization} that facilitates the transfer of feature knowledge related to the global context from the non-streaming Transformer to the tailored streaming Transformer through weight inheritance. Further, we propose a cross-modal structure based on knowledge distillation (KD), named \textit{cross-modal KD}, to mitigate the negative effect of low quality mmWave signals on recognition performance. In the \textit{cross-modal KD}, the audio streaming Transformer provides feature and response guidance that inherit fruitful and accurate speech information to supervise the training of the tailored radio streaming Transformer. The experimental results show that our Radio2Text can achieve a character error rate of 5.7\% and a word error rate of 9.4\% for the recognition of a vocabulary consisting of over 13,000 words.

\end{abstract}




\maketitle

\section{Introduction}
With the rapid progress in wireless sensing, one of the sensing modalities, radio frequency (RF) signals, have gained significant attention in academia and industry, as their ability to work under various lighting conditions without raising privacy concerns and to traverse the occlusions \cite{surveywireless,surveymmwave}. The unique properties of RF signals enable them to facilitate various applications, from sensing human physical and physiological activities \cite{heartbeat,RFWash,end2endradar,humanpose,rf3d} to detecting object vibration \cite{mmvib}. 
Among many RF techniques, millimeter wave (mmWave) signals have the distinctive ability to detect subtle displacements with high accuracy, even the vibration of the sound source at the millimeter level, owing to their short wavelength. Moreover, unlike conventional solutions using microphones, mmWave signals can resist noise or irrelevant sound and penetrate soundproof obstacles \cite{waveear,wavesdropper}. Therefore, mmWave signals can be used not only to perceive audio but also to extend audio-related applications to noisy and soundproof scenarios, where microphones fail.  

As a way that can directly acquire content information, speech recognition is the cornerstone for various audio-related applications \cite{asrreview}. mmWave empowers speech recognition to achieve previously unattainable applications, such as conference speech transcription and eavesdropping in noisy and soundproof scenarios. In real applications, reducing latency is of central importance in improving efficiency and user experience, and the size of recognizable vocabulary directly influences the effectiveness of the recognition system. Thus, it is desirable to achieve streaming automatic speech recognition (ASR) on text with a large vocabulary, rather than waiting for the completion of a full speech utterance before recognizing the entire sentence and only performing recognition within a restricted vocabulary size. This expectation should also be followed for mmWave-based speech recognition. 

\begin{figure}[t]
  \centering
  \includegraphics[width=4.7in]{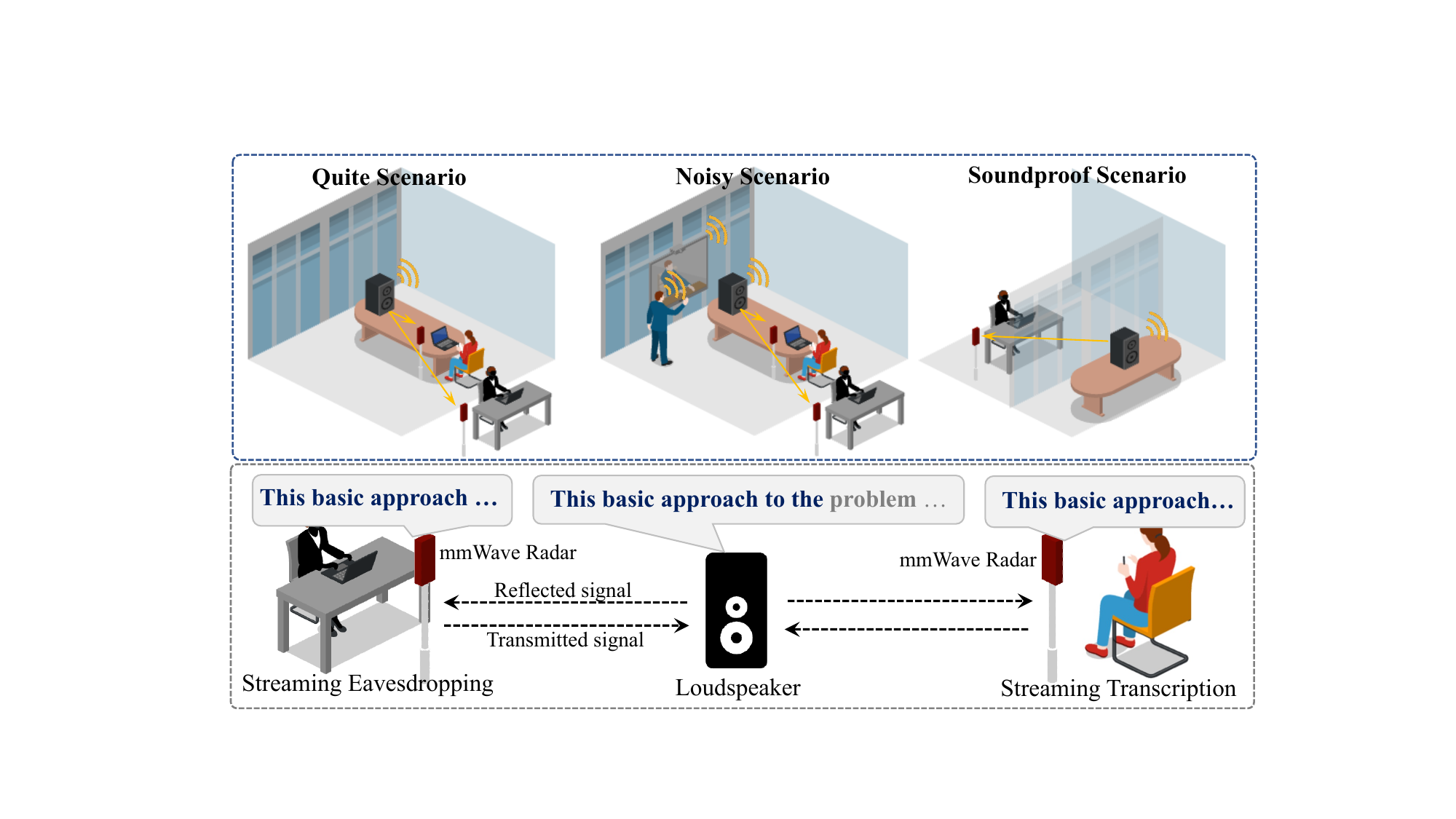}
  \caption{Illustration of Radio2Text working scenarios.}
  \label{overview}
\end{figure}

However, existing solutions neglect such potential requirements. They either utilize pre-recovered speech from mmWave signals for further speech recognition or recognize text from mmWave signals via non-streaming models. Also, the number of words they can recognize is restricted to within the range of one thousand.
For example, WaveEar \cite{waveear} and mmEavesdropper \cite{mmEavesdropper} use mmWave signals to recover speech from the vibration of the human throat and loudspeaker and then recognize the text with 814 and 36 distinct words from the recovered speech. AmbiEar \cite{ambiear} and Wavesdropper \cite{wavesdropper} directly recognize the text with 30 and 57 words from mmWave signals of surrounding object's vibration and vocal response in a non-streaming mode. The absence of streaming capability and the constraint on the number of recognizable words limit their application in the real world. 

In this paper, we introduce \textit{Radio2Text}, a system that uses mmWave signals for streaming speech recognition. \textbf{\textit{Radio2Text} can accurately recognize the text containing a vocabulary of over 1,3000 words in a low-latency streaming mode without recovering speech in advance, even in noisy and soundproof scenarios.} For the potential usage scenarios of \textit{Radio2Text} (i.e., conference speech transcription and eavesdropping), loudspeakers are indispensable components since they are widely deployed in people's daily life as the information broadcaster of communication and conference systems. For example, smart devices leverage loudspeakers to replay users' commands or private information, and online meetings use loudspeakers to play conversations of remote participants. Therefore, our work focuses on loudspeakers as transcription and eavesdropping targets. Figure \ref{overview} shows the typical working scenarios of Radio2Text. mmWave radar transmits signals to the loudspeaker and parses the reflected mmWave signals for streaming ASR, i.e., contents are recognized at the same time as the loudspeaker plays. It can be used to transcript speech in online meetings from the loudspeaker, as well as to eavesdrop on the speech from the loudspeaker when people are meeting or communicating, even in the presence of noise and soundproof materials. Notably, Radio2Text has the capability to recognize text with a large vocabulary of over thirteen thousand words instead of basic commands or sentences with a limited vocabulary.

To achieve our goal of implementing a mmWave-based system that can perform streaming ASR on large vocabulary text, we need to address two challenges. 
The first challenge arises from the aspect of the ASR network. Recognizing over ten thousand words is obviously challenging, especially with low quality mmWave signals (detailed below). There may be more potential matches for a given inaccurate input, which increases the possibility of ambiguity in recognition. This requires a network with powerful feature representation ability to effectively extract acoustic features from low quality mmWave signals and establish correlations with limited-seen features. Moreover, to achieve streaming speech recognition, ASR models can only rely on historical information (i.e., features prior to the current input) for prediction without incorporating entire future information. Thus, streaming ASR models can not use the global context to enhance feature interaction and extraction, resulting in inferior performance compared to non-streaming models. 

The second challenge is associated with the inherent attributes of mmWave signals. Restricted by the design of radar, the upper frequency that radar can perceive is limited, i.e., below 1.5 KHz. The missing high-frequency band compromises the integrity of speech information. Moreover, reflected signals are likely to be flooded by the noise. This is because the vibration amplitude is inherently weak. Further, the reflected mmWave signal attenuates as it propagates and is susceptible to noise from the wireless channel and multipath signals. These factors result in a degradation in the quality of speech-related mmWave signals, which further compromises the performance of the streaming ASR network. The signal processing methods for denoising can only mitigate some of the negative effects of noise, which provides limited help in improving the recognition performance of streaming ASR networks for text with a large vocabulary. 

The design of \textit{Radio2Text} takes into account these challenges and tackles them as follows:

To address the first challenge, we incorporate the encoder-decoder-based Transformer with the chunk-wise setting and the triggered attention, to construct a powerful streaming ASR network, named tailored streaming Transformer. The incorporation enables the network to effectively extract acoustic representations and adaptively construct correlations between contexts from limited-seen features, while ensuring streaming attributes. Moreover, the inability of streaming networks to utilize entire future information degrades their performance. To bridge the performance gap between the streaming model and the non-streaming model, we propose a method called \textit{Guidance Initialization (GI)}, which uses the trained parameters of the selected layers of the trained non-streaming model to initialize the corresponding layer of the streaming model. This allows the streaming model to inherit the feature knowledge that is relevant to the global context from the non-streaming model.

For the second challenge, the low quality mmWave signals result in suboptimal performance when they are directly fed into the tailored streaming Transformer. Fortunately, the consistency between the information of speech-related mmWave signals and paired speech signals provides a new angle to tack this issue. They share the same information representation and target. Also, the paired speech signals contain abundant high-frequency components and are not interfered by noise, thus providing sufficient acoustic features. Inspired by this, we propose a cross-modal structure based on knowledge distillation (KD), named \emph{cross-modal KD}, to transfer the knowledge from audio modality to radio modality. The audio streaming Transformer is used as the teacher model to provide the features and responses that inherit fruitful and accurate speech information as guidance. Then, the knowledge of the teacher model is distilled to the student model, i.e., radio streaming Transformer, through the distillation loss of feature-based and response-based knowledge for the streaming Transformer. Once our tailored streaming Transformer is trained through the \emph{cross-modal KD}, it only uses mmWave signals as input for inference. As a result, the tailored streaming Transformer is capable of recognizing text accurately from low quality mmWave signals, guided by learned knowledge from the audio domain.

Overall, our contributions are summarized as follows:
\begin{itemize}
    \item We first introduce the concept of streaming ASR for mmWave signals and extend the recognition vocabulary size to enable low-latency applications that recognize a wider range of words in practice. To realize it, we design a mmWave-based streaming ASR system \textit{Radio2Text}, which recognizes the text with a large vocabulary of over 1,3000 words in real-time, without waiting for the loudspeaker to play the entire speech.
    
    \item On the basis of the tailored streaming Transformer that is powerful for feature representation, we propose the \emph{Guidance Initialization} to inherit the feature knowledge of global context from the non-streaming ASR network and propose the \emph{cross-modal KD} to transfer the knowledge from audio modality to radio modality in feature and response levels. This enables the tailored streaming Transformer to accurately recognize the text with a large vocabulary from low quality mmWave signals.
    
    \item We implement the Radio2Text on the Commercial off-the-shelf (COTS) mmWave radar and evaluate it with extensive experiments, which shows that \emph{Radio2Text} achieves a character error rate of 5.7\% and a word error rate of 9.4\% for a vocabulary consisting of over thirteen thousand words, and \emph{Radio2Text} outperforms compared mmWave-based methods in quiet, noisy and soundproof scenarios.  
\end{itemize}

\section{Preliminary}
\subsection{mmWave-based Sensing} \label{mmwavepreliminary}
Our mmWave-based streaming ASR relies on the mmWave frequency-modulated continuous wave (FMCW) radar, which is widely used in autonomous driving \cite{fmcwautonomous} and industry \cite{jardak2019compact}. 
Specifically, FMCW radar generates a periodic chirp signal and transmits it to the target object, and the received signal reflected from the target carries the propagation distance and vibration information. 
Assume that the distance between the target and radar is $R(t)=R_0+x(t)$, where $R_0$ is the fixed distance and $x(t)$ is the vibration displacement. Denoting the transmitted signal $s(t) = \exp[j(2\pi f_{c}t+\pi Kt^2)]$, the received signal with a round trip delay of $t_d=2R(t)/c$ can be expressed as
$r(t) = As(t-t_d) = A\exp[j(2\pi f_{c}(t-t_d)+\pi K(t-t_d)^2)]$, where $A$ is the propagation loss, $f_c$ is the carrier frequency and $K$ is the chirp slop of FMCW signal. Then a mixer is used to demodulate the received signal and filter out high-frequency carrier wave, thereby generating the intermediate frequency (IF) signals
\begin{equation}
    y(t) = s^*(t)r(t)
    \approx A\exp(j2\pi K{t_d}t+j2\pi f_{c}t_d).
\end{equation}

It can be seen that frequency and phase of IF signals are related to round trip delay that includes vibration signals. However, the frequency changes induced by vibration during a chirp period are relatively small compared to the observation window (i.e., observation period in Fourier Transforms), so it is not discernible in the frequency spectrum. To obtain vibration signals, we need to extract the phase differences across consecutive chirps. 
We should first perform FFT on the reflected signal within a chirp, known as range FFT, to find the signal that corresponds to the distance value at which the expected object is placed. We then combine the samples in the selected distance value (i.e., range bin) cross chirps into a sample sequence, which can be expressed as: 
\begin{equation}
    Y(t)=A\exp[j4\pi f_cR(t)/c].
\end{equation}
The sample rate of mmWave signals can be considered as the chirp rate of mmWave radar (i.e., chirps per second in this paper), which covers the frequency range of human speech. 

Loudspeakers modulate speech signals and express them in the form of vibration signals. The radar transmits mmWave signals to the target loudspeaker vibrating as $x(t)$, and the reflected signals $r(t)$ carry the vibration information related to the speech. 
A simple way to obtain speech-related vibration signals $x(t)$ is to extract phase changes $\phi_n$ (\emph{n}th sample) across multiple chirps from the above signal $Y(t)$ and unwrap the phase changes as $x(n)=unwrap(\phi_n)\cdot c/4\pi f_c - R_0$.  

\subsection{Encoder-Decoder based Transformer} \label{transformer}
Transformer achieves impressive results in speech recognition due to its ability to capture long-term dependencies \cite{transformer}. It follows an attention-based encoder-decoder (AED) structure, where the encoder extracts a sequence representation from the input sequence and the decoder generates an output sequence in an auto-aggressive mode from the encoder representation.
Specifically, the encoder consists of several identical layers, each of which contains a multi-head self-attention (MHSA) layer and a feed-forward network (FFN), connected by residual connections and layer normalization (LN). The only difference with the encoder is that the decoder inserts an additional MHSA layer between the two sub-layers in each encoder layer as the cross-attention layer to interact with the encoder output. Especially the MHSA layer is an ensemble of multiple parallel self-attention layers, which can be described as mapping query and a set of key-value pairs to an output. Suppose the input of the Transformer block is linearly transformed to $Q, K$ and $V$, the output of MHSA is calculated by:
\begin{equation} \label{mhsa}
\begin{aligned}
    & MultiHead(Q,K,V)=Concat(head_1,...,head_H)W^{head}, \\
    & head_h=Attention(QW_h^Q, KW_h^K, VW_h^V)=softmax(\frac{QW_h^Q(KW_h^K)^T}{\sqrt{d_k}})VW_h^V, \\
\end{aligned}
\end{equation}
where $W_h^Q, W_h^K, W_h^V$ and $W^{head}$ are trainable projection matrices for queries, keys, values and outputs, and $1/\sqrt{d_k}$ is the scaling factor. In the self-attention layer of the encoder, keys, values and queries originate from the same input, and every position of input attends to all other positions. Compared with the encoder, each position in the self-attention layer of the decoder can only attend to the past information (i.e., the position before the current position) due to the auto-regressive property. While in the intermediate cross-attention layer of the decoder, the queries come from the decoder and the keys and values come from the encoder, and thus every position in the decoder attends over all positions in the encoder. The self-attention on the full input sequence violates the principle of streaming ASR, which only considers past information and disregards entire future information. Therefore, traditional Transformers cannot be used for streaming ASR without special designs.

\begin{figure}[t]
  \centering
    \subfigure[Waveforms]{
    \label{Waveforms}
    \includegraphics[width=1.4in]{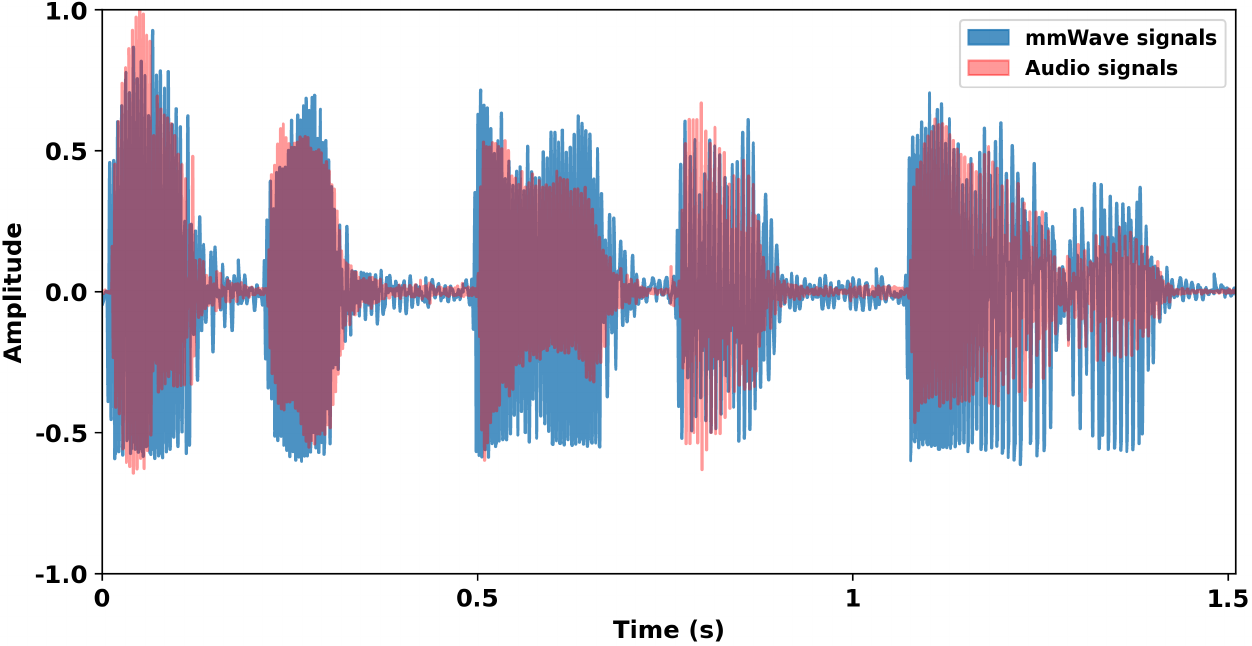}} 
    \hfil
    \subfigure[Mel-Spectrograms]{
    \label{melspectrograms}
    \includegraphics[width=1.82in]{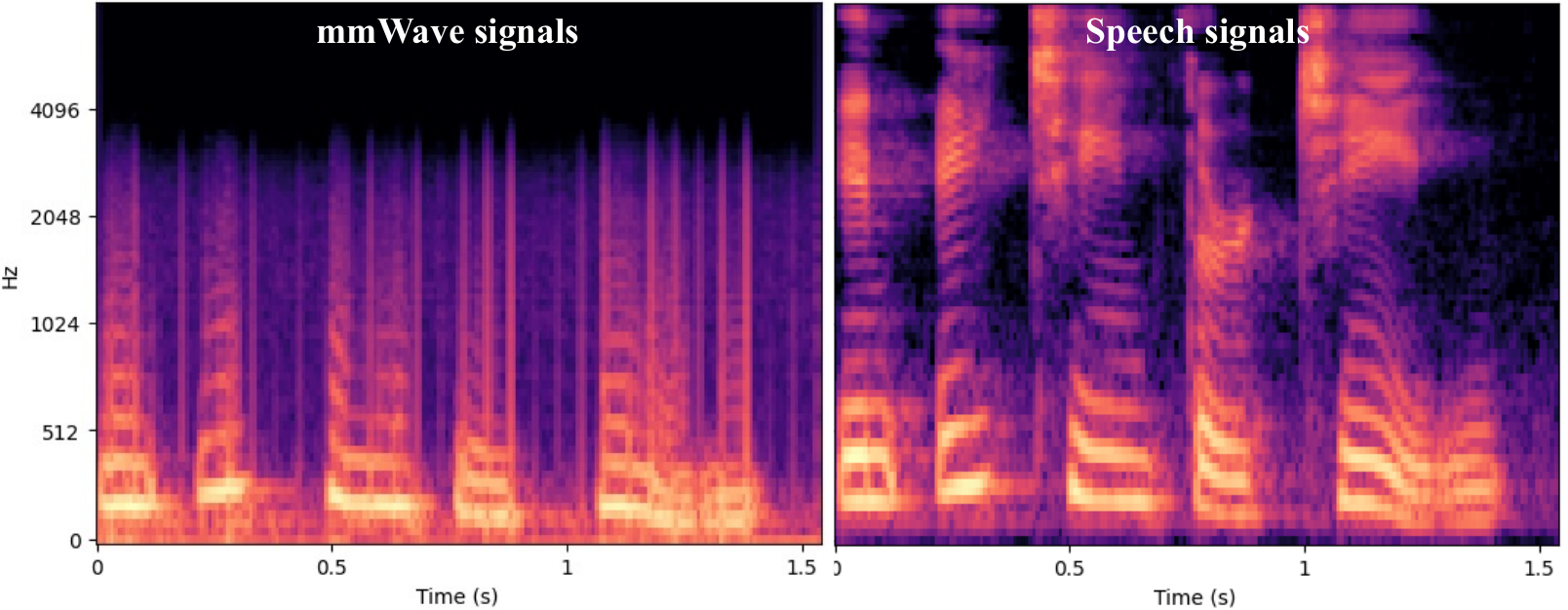}} 
    \hfil
    \subfigure[Cross-correlation]{
    \label{Crosscorrelation}
    \includegraphics[width=1.5in]{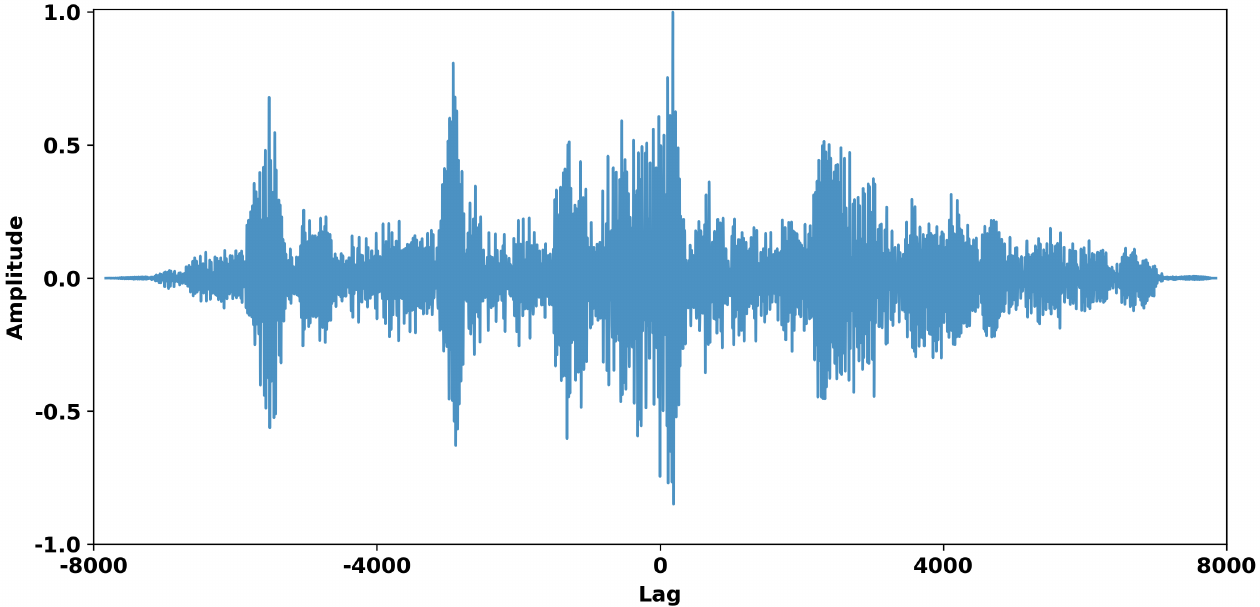}} 
    \hfil
    \subfigure[ASR error rate]{
    \label{errorrate}
    \includegraphics[width=1.2in]{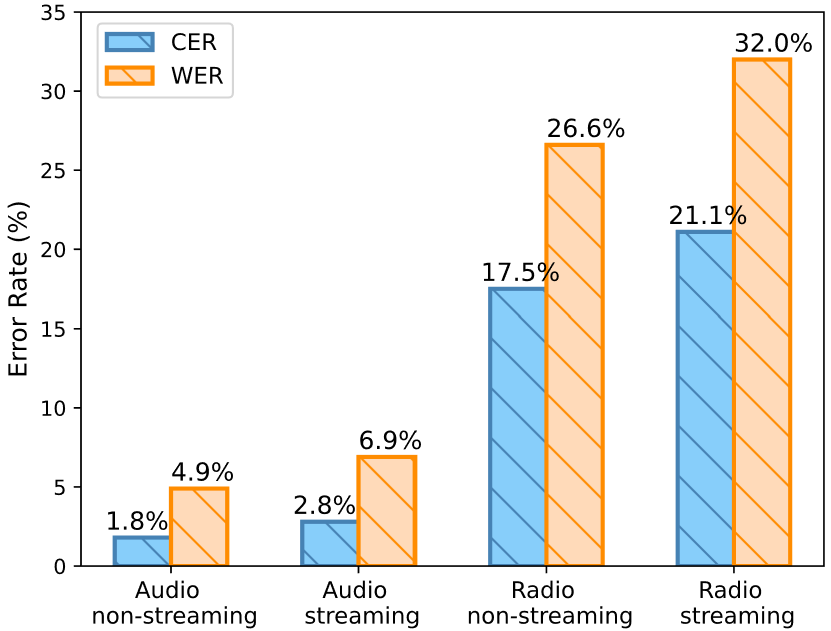}}
    \hfil
  \caption{Feasibility study results: the correlation between the original speech signals and the demodulated mmWave signals, and the error rate of speech recognition for streaming and non-streaming models.}
  \label{preliminarystudy}
\end{figure}

\section{Feasibility Study and Challenges} \label{feasibilitystudy}
We conduct some preliminary experiments to investigate the feasibility and challenges of detecting the vibration of loudspeakers using mmWave radar and recognizing text from mmWave signals using a streaming ASR network.  

\subsection{Correlation Analysis} \label{correlationanalysis}
Existing networks for speech recognition rely on audio signals as the input. Thus, when we adapt the network to the mmWave domain, the correlation between the original speech signals and demodulated mmWave signals needs to be analyzed. 
We conduct a preliminary experiment by using a COTS mmWave radar to sense the front loudspeaker playing a simple utterance (i.e., IN EIGHTEEN THIRTEEN).
As shown in Figure \ref{Waveforms}, waveforms of original speech signals and demodulated mmWave signals share similar trends in the envelopes and exhibit fluctuations at comparable positions. 
It can be seen in Figure \ref{melspectrograms} that demodulated mmWave signals and original speech signals remain consistent in the low-frequency band, which plays a dominant role in the frequency range of human speech. We further quantitatively evaluate their similarity through cross-correlation \cite{crosscorrelation}, which calculates the displacement of one sequence relative to the other to reflect the similarity. For two time-aligned sequences, there should be a peak at a lag of zero if they are similar. The results in Figure \ref{Crosscorrelation} are consistent with the description, indicating a significant correlation between mmWave signals and speech signals. 
The observation demonstrates the feasibility of using mmWave signals to recognize speech content.

\subsection{Streaming Analysis} \label{streaminganalysis}
The definition of streaming ASR is to recognize each word in real-time as the input speech utterance is processed, instead of waiting for the completion of a full speech utterance before recognizing the entire sentence. To satisfy the requirement, the self-attention and cross-attention mechanisms in the streaming Transformer need to be constrained to only attend to past information without entire future information. This compromises the traditional Transformer's ability to extract global context, causing performance degradation. We use an audio dataset and a radio dataset to evaluate the performance of the streaming Transformer and the non-streaming Transformer, respectively. 
As shown in Figure \ref{errorrate}, compared with the non-streaming Transformer, there is an obvious increase in the error rate of the streaming Transformer in both audio and radio datasets.

\subsection{mmWave Signals Analysis} \label{mmwavesignalsanalysis}
Although the demodulated mmWave signal exhibits a strong correlation with the original speech signal, the quality of mmWave signals is inferior to that of speech signals. In addition to the similarity, we can also derive from Figure \ref{melspectrograms}: (1) The upper frequency that mmWave radar perceive is limited, normally below 1.5 KHz. 
Compared with speech signals, mmWave signals lack abundant high-frequency information, which compromises the integrity of speech information; (2) mmWave signals contain a significant amount of noise, and particularly the portion that lacks high-frequency information is even drowned out by noise. 
Therefore, the limited frequency range and the presence of noise significantly reduce the quality of demodulated mmWave signals. As shown in Figure \ref{errorrate}, although the streaming network achieves impressive performance in the audio domain, directly feeding raw mmWave signals into the streaming ASR Transformer results in a drastic drop in streaming recognition results.

\section{System overview}
\begin{figure}[t]
  \centering
  \includegraphics[width=5.6in]{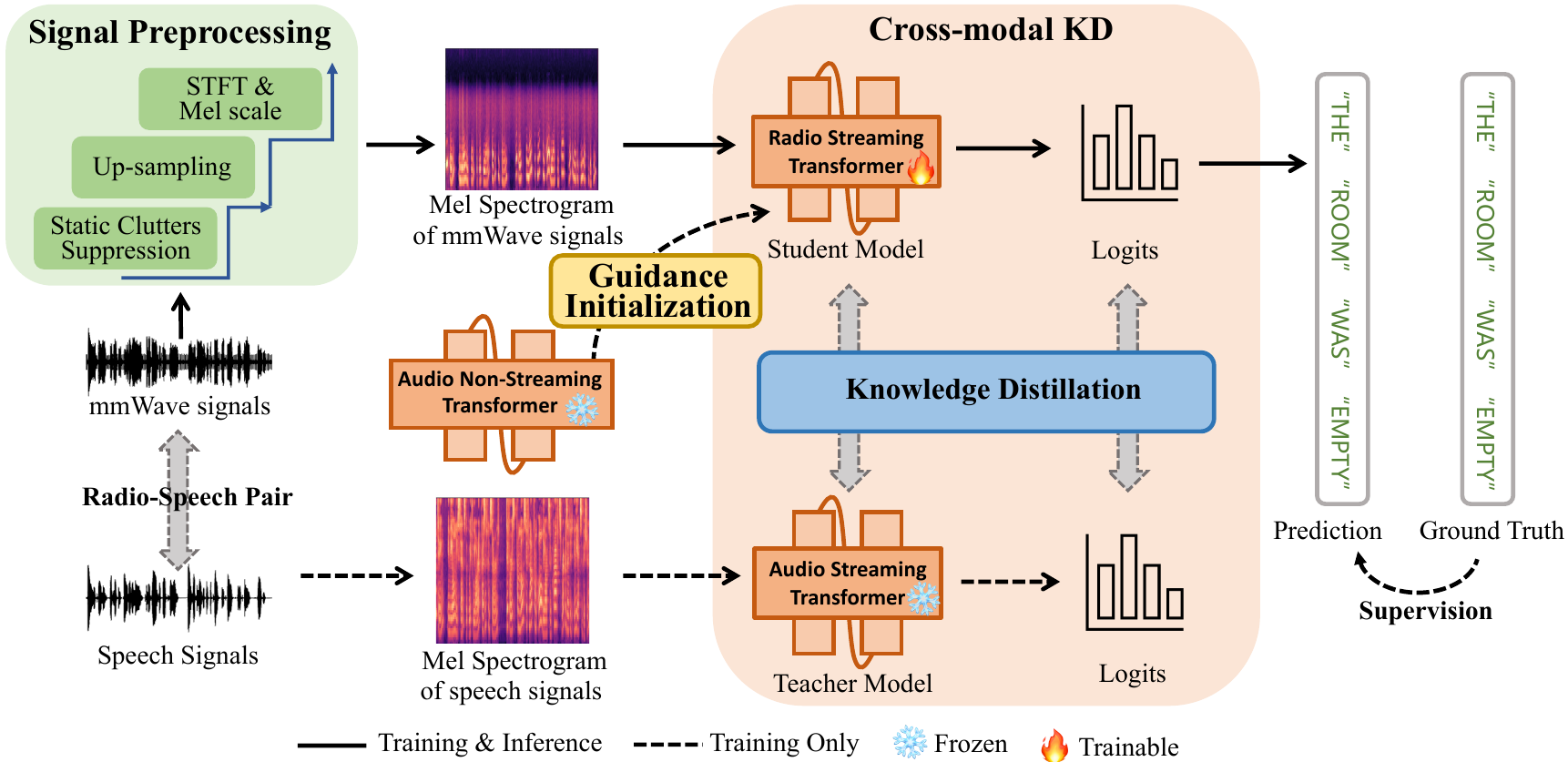}
  \caption{The system overview of Radio2Text. Dotted lines indicate training only, and solid lines represent training and inference. Snowflake represents frozen network without parameter updating, and flame represents a trainable network.}
  \label{framework}
\end{figure}

The proposed Radio2Text aims to accurately recognize the text with a large vocabulary in real-time from low quality mmWave signals, extending mmWave-based ASR systems to realistic scenarios. Figure \ref{framework} depicts the overall structure of the proposed Radio2Text.

In the training stage, the collected speech-related mmWave signals and the corresponding speech signals are used as input pairs. After signal preprocessing, mmWave signals are transformed into a Mel-Spectrogram with the same sample rate as speech signals. The paired speech signals are also converted into Mel-Spectrogram. Then, the radio streaming Transformer is initialized by the trained weights from the audio non-streaming Transformer through the proposed GI. Subsequently, in the cross-modal KD, the audio streaming Transformer, as the teacher model, infers only form input speech signals to provide features with accurate and fruitful speech information as the feature guidance and provide the prediction logits as the response guidance. The radio streaming Transformer is the trainable student network for generating speech recognition results from mmWave signals. The student leverages the guidance in feature and response levels from the teacher for cross-modal knowledge transfer in training. It also accepts guidance from the ground truth label for supervised training. 
In the inference stage, the student model, radio streaming Transformer, is used alone to generate the recognition results from mmWave signals processed by signal preprocessing. To achieve the goals, Radio2Text integrates four components:

\textbf{Signal Preprocessing.} \label{preprocessing}
We propose a signal preprocessing scheme, which aims to suppress the static clutter in speech-related mmWave signals, and then generate the Mel-Spectrogram of mmWava signals with the same frequency as the paired speech signals, enabling knowledge transfer in the proposed cross-modal framework. 

\textbf{Tailored Streaming Transformer.} The goal of this component is to implement streaming recognition from input mmWave signals. The tailored streaming Transformer is powerful in feature representation, providing the basis for real-time recognition of text with a large vocabulary. 

\textbf{Guidance Initialization.} This component aims to alleviate the deficiency of the tailored streaming network unable to access entire future inputs. We propose the GI, which uses the trained weights of the selected layers of the non-streaming Transformer as guidance to initialize the tailored streaming Transformer, thereby inheriting the global context related feature knowledge of the non-streaming Transformer.

\textbf{Cross-modal Knowledge Distillation.} To mitigate the negative effect of low quality mmWave signals, we design the cross-modal KD. The well-trained audio streaming Transformer provides features with fruitful speech information and logits of accurate prediction for guidance. Under such guidance, the trainable radio streaming Transformer effectively learns the latent information concealed by the noise and the class probability distribution, enabling accurate recognition from low quality mmWave signals. 

\section{System Design}
\subsection{Signal Preprocessing}
The demodulated mmWave signals contain static clutters generated from surrounding static objects, which can cause signal deviation in the I/Q domain and degrade signal quality \cite{mmvib}. To eliminate static clutters, we first fit a circle based on the demodulated mmWave signals in the I/Q domain and then translate the circle center to the origin of coordinates. Moreover, the vibration signals are incompatible with the paired input speech signals in frequency. Therefore, we upsample the vibration signals into the same frequency as the paired speech signals using the cubic spline interpolation \cite{interpolation}.  Subsequently, we apply Short-Time Fourier Transform (STFT) on mmWave signals to calculate the spectrogram and map it to Mel scale to produce Mel-Spectrogram. This is because the Mel scale, a non-linear scale, is more in line with human perception of audio frequency. Also, the Mel-Spectrogram provides a representation of changes in frequency over time, which is a highly information-dense representation of speech signals compared with raw time series.

\subsection{Streaming Transformer} \label{streamingtransformer}
Streaming ASR imposes constraints on the range of accessible features for prediction, and recognition of text with a large vocabulary increases the possibility of ambiguity in matches between features and potential text, especially for low-quality radio signals. Given these challenges, the backbone network possessing powerful feature representation ability is the basis for streaming ASR with a large vocabulary. In this paper, we leverage the encoder-decoder-based Transformer as the backbone for streaming ASR. This is because Transformer can effectively extract features and adaptively construct correlations between contexts \cite{transformer}, even at the compromise of only accessing limited context for streaming settings. Moreover, the AED structure has the ability to capture dependencies between inputs and outputs to benefit feature representation \cite{streamingaed}. Compared with the alternative streaming structure of RNN-T (Recurrent Neural Network Transducer) \cite{rnnt}, AED presents advantages in terms of generalization to smaller data sets and lower memory demand for training \cite{rnnttraining}. Although the encoder-decoder-based Transformer compromises the range of accessible context for streaming, it is still powerful in learning representations of speech-related features and establishing correlations among limited-seen features.

\begin{figure}[t]
  \centering
  \includegraphics[width=5in]{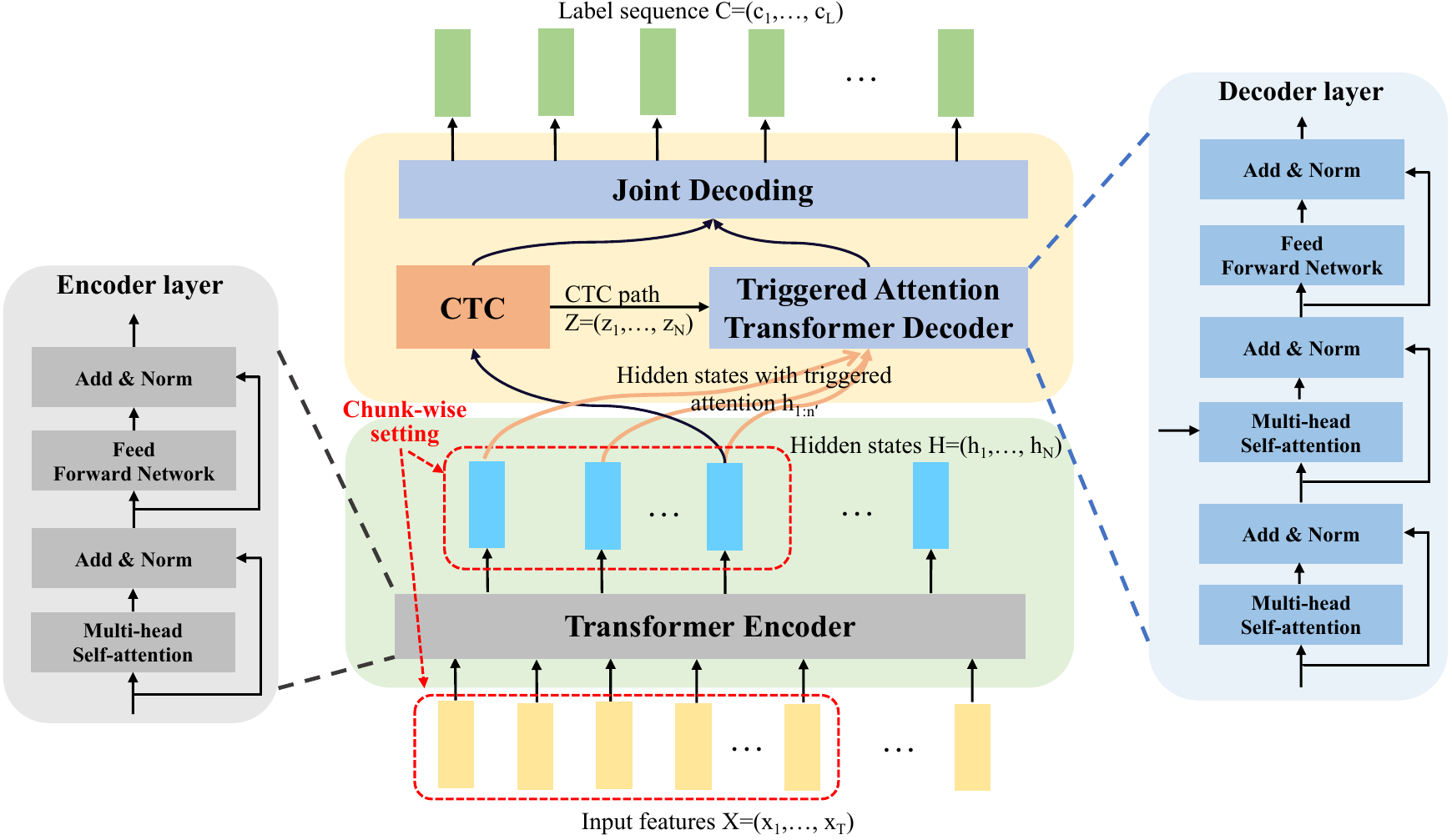}
  \caption{The architecture of the tailored streaming Transformer.}
  \label{taframework}
\end{figure}

As mentioned in Section \ref{transformer}, encoder-decoder-based Transformer networks are incapable of supporting streaming speech recognition in nature, because self-attention in the encoder and cross-attention in the decoder require an entire input sequence to calculate attention. To address this issue, we extend chunk-wise setting \cite{chunksetting} and triggered attention \cite{tastreamingtransformer} to the encoder-decoder-based Transformer, named the tailored streaming Transformer. Specifically, we use the chunk-wise setting to flexibly control the attention range in the self-attention layers of the encoder, while we employ triggered attention to decide the activation of cross-attention in a decoder. Although some efficient Transformers also do not attend all positions in self-attention, they mainly aim to sparsify the attention matrix by limiting the field of view to be fixed without especially considering restricting the size of accessible future positions \cite{efficienttransformer}. Therefore, most efficient Transformers are inapplicable for the streaming mode. Figure. \ref{taframework} shows the overall structure of the streaming Transformer, which mainly includes two components: chunk-based encoder, and triggered-attention-based decoder. The encoder converts the input mmWave features $X=(x_1,...,x_T)$ into the hidden states $H=(h_1,...,h_N)$, which is shared by connectionist temporal classification (CTC) module and triggered-attention-based decoder to predict the label sequence $C=(c_1,...,c_L)$, where $T$, $N$ and $L$ represent the length of features, hidden states and labels.

\subsubsection{Chunk-based Encoder} Chunk-based encoder consists of a VGG-like convolution module and a stack of Transformer encoder layers with the chunk-wise setting. We utilize the VGG network to reduce the frame rate of features and extract the local features of input speech-related mmWave signals in lieu of the position embedding in the Transformer. This is because the local features, such as phoneme transitions and pitch variations, are critical for speech recognition than simple positional embeddings \cite{position}, such as sine and cosine functions. 
As described in Section \ref{transformer}, each encoder layer is composed of the MHSA and FFN connected by the residual layer and LN, which can extract global features to capture the relation between different characters. The chunk-wise setting is a crucial component of the streaming Transformer's encoder, as it segments the input sequence into fix size chunks, enabling self-attention to be applied in a streaming fashion. This setting is a masking strategy that generates a binary mask $m_h$ to control the self-attention scores in the $softmax$ function of Eq. \ref{mhsa}. 
As shown in Figure \ref{taframework}, the frames within the same chunk can see each other, and the frames in the current chunk can also see other frames in the previous chunks, enabling attention to history information and limited future information. While maintaining the streaming mode, this strategy helps achieve high accuracy by allowing for future information. 

\subsubsection{Triggered-Attention-based Decoder} The decoder with triggered attention is a hybrid of Transformer decoder layers and triggered attention mechanism. The former is responsible for converting the hidden states of the encoder into the corresponding transcription, while the latter is used to enable the decoder to operate in a streaming mode. The triggered attention is based on a CTC model to condition the attention mechanism of the decoder only on chunks that belong to past encoder frames. Specifically, it aligns the hidden states of the encoder and label sequence, and further determines the index of hidden states corresponding to each label (named trigger events). 
The alignment is provided by a CTC model, which forces alignment between the variable-length hidden states and the transcriptions through an auxiliary CTC objective $p_{ctc}(C|H)$ \cite{ctc}.
Then, the CTC model output (CTC path) $Z$ of the highest overall probability is converted into the trigger event sequence $Z^{'}$ by only keeping the first occurrence of each label and replacing other labels with the blank symbol $\phi$. 
For example, the CTC path $Z=(\phi, C, C, A, A, \phi, T, \phi)$ generated from the feature sequence with the label $C=(C,A,T)$ is converted to the trigger event sequence $Z^{'}=(\phi, C, \phi, A, \phi, \phi, T, \phi)$. The triggering event represents the index of the hidden state required for generating each label.
Thus, the triggered attention based decoder is capable of predicting the label based only on hidden states prior to the corresponding index without relying on future states, thus operating in a streaming mode. 
To match the chunk-wise setting, we use the index of the chunk that the trigger event belongs to as the frame index of the trigger event.
Correspondingly, the object of triggered attention can be written as $P_{ta}(C|H)=\prod_{l=1}^{L}P(c_l|c_{1:l-1},h_{1:n^{'}})$, where $n^{'}$ denotes to the last frame index within the chunk where the trigger event $Z{'}$, corresponding to the label $c_l$, resides, and $P(c_l|c_{1:l-1},h_{1:n^{'}})$ represents the output of Transformer decoder.

\subsubsection{Training and Decoding} Following a widely used training strategy of hybrid CTC/Attention training \cite{jointctcattention}, the tailored streaming Transformer is jointly trained by the CTC loss and attention loss in the following manner: 
\begin{equation}
    \label{hybrid}
    \mathcal{L}_{hybrid}=-\lambda\log P_{ctc}-(1-\lambda)\log P_{ta},
\end{equation}
where $\lambda$ is a tunable parameter. With regard to decoding, we use the CTC-triggered attention decoding for joint scoring of the CTC and transformer model outputs \cite{tastreamingtransformer}. The decoding algorithm is based on the frame-synchronous prefix beam search algorithm \cite{prefixbeamsearch}, extending it through the integration of the triggered-attention-based decoder.

\begin{figure}[t]
  \centering
  \includegraphics[width=3in]{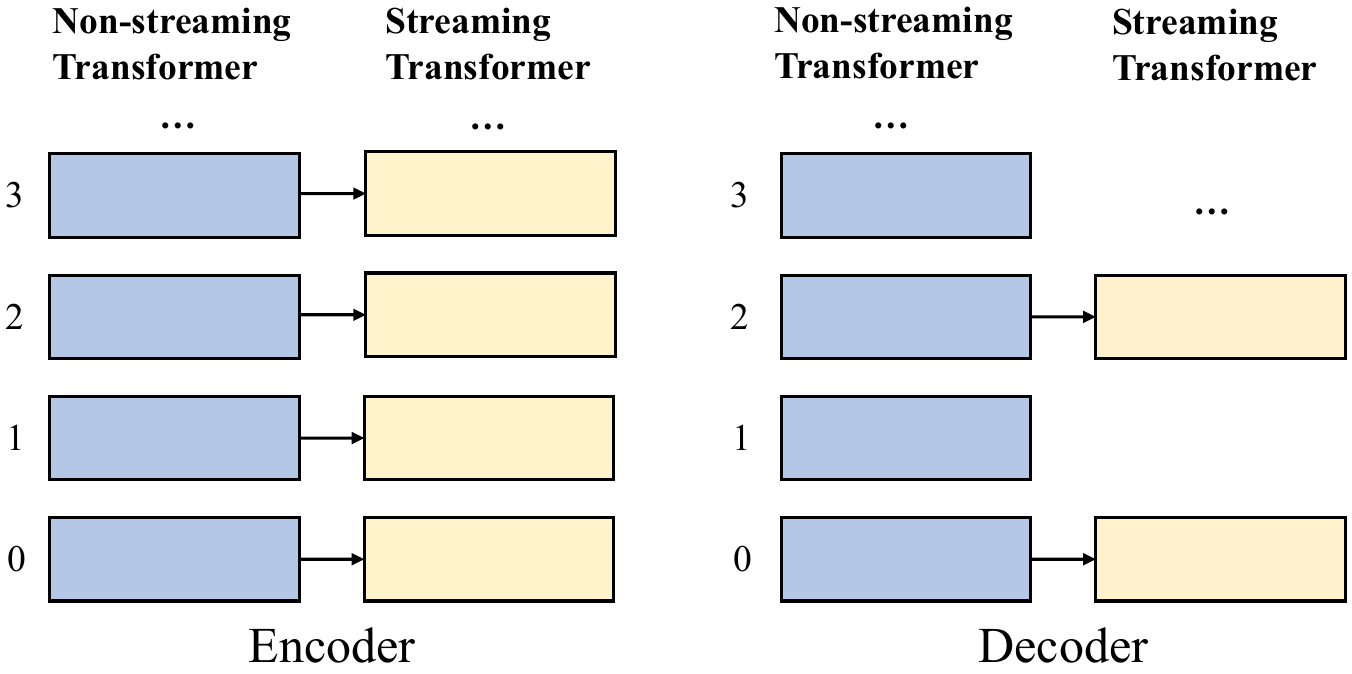}
  \caption{Matching strategy of Guidance Initialization.}
  \label{matchinggi}
\end{figure}

\subsection{Guidance Initialization} \label{tgi}
The inability to capture global context due to the limitation of the streaming setting makes the streaming Transformer to be inferior to the non-streaming Transformer. The intuition is to leverage KD to distill the knowledge from the non-streaming Transformer to the streaming Transformer. However, this way causes a conflict with the proposed cross-modal KD (detailed below), as these two different teachers provide distinct feature guidance about future information. Considering the trained network's weights determine the feature map value, acquiring knowledge from the weights of the non-streaming Transformer is an alternative to providing feature guidance to the streaming Transformer. Therefore, we propose GI that uses the trained weights of the selected layers of the non-streaming Transformer to initialize the corresponding layers of the streaming Transformer. The streaming Transformer can inherit the feature knowledge that is relevant to the global context from the non-streaming Transformer. Based on the inherited weights, the streaming Transformer can collaborate well with cross-modal KD to further enhance feature extraction ability.

In our design, we use the audio non-streaming Transformer as the trained weight provider and our radio tailored streaming Transformer as the acceptor. Also, the depth of the non-streaming Transformer is equal to that of the streaming Transformer. Therefore, we directly match the non-streaming Transformer and streaming Transformer layer by layer for the selected pair of layers. Since the representations learned in different layers vary a lot \cite{layersimilarity}, it is vital to select the proper pair of layers to match. In the encoder-decoder structure, the encoder layer plays the role of feature extraction, and the decoder is mainly used for establishing the relation between extracted features and prediction. Therefore, to inherit the feature knowledge related to global context from the non-streaming Transformer, we select all the encoder layers to inherit the trained weights. Also, to enhance the cooperation between the inherited feature knowledge and the decoder prediction, we evenly select the decoder layer for weight inheritance. The matching strategy is shown in Figure \ref{matchinggi}. We assume the non-streaming Transformer has $J$ encoder layers and $K$ decoder layers, and the parameters of each layer can be represented as $(W_{NT-E}^1, ..., W_{NT-E}^J, W_{NT-D}^1, ..., W_{NT-D}^K)$. Similarly, the parameters of each layer in the streaming Transformer are $(W_{ST-E}^1, ..., W_{ST-E}^J, W_{ST-D}^1, ..., W_{ST-D}^K)$. The matching strategy for the $j$-th encoder layer and $k$-th encoder layer can be formulated as
\begin{equation}
\begin{aligned}
    &W_{ST-E}^j=W_{NT-E}^{j} , j=0,1,2, ..., J \\
    &W_{ST-D}^k=W_{NT-D}^{k} , k=0,2,4, ..., K. \\
\end{aligned}
\end{equation}

\subsection{Cross-modal Knowledge Distillation} \label{crossmodal}
The absence of high-frequency bands and the presence of significant amounts of noise results in inferior quality of mmWave signals. Thus, it is challenging for networks to learn useful acoustic features for recognition, resulting in a decreased performance in streaming speech recognition, as discussed in Section \ref{mmwavesignalsanalysis}. 
The paired speech signals share the same information representation and target with mmWave signals. Also, the paired speech signals contain abundant high-frequency components and are not interfered by noise, thus providing sufficient and accurate acoustic features. Inspired by this, we propose the cross-modal kD, where the trained audio network serves as the teacher and the mmWave network is the student, to transfer the knowledge through the teacher-student training paradigm. The cross-modal KD uses the knowledge provided by the trained audio network in feature and response levels to guide the radio network. The feature guidance enables the radio network to imitate high-quality feature representations of the audio network, facilitating the learning to complement absent high-frequency components and to extract clean features from input mmWave signals that lack high-frequency bands and contain significant noise. Also, the radio network can mimic the soft logits of the audio network with response guidance, fitting accurate predictions in the presence of low-quality input. In our design, we follow the common KD paradigm where the teacher and student have homogeneous settings. Thus, the teacher and student use the same network structure, named audio streaming Transformer and radio streaming Transformer, so that feature representations can be directly transferred without pruning.

\begin{figure}[t]
  \centering
    \subfigure[Encoder knowledge distillation for hidden states and attention matrices]{
    \label{encoderkd}
    \includegraphics[width=1.9in]{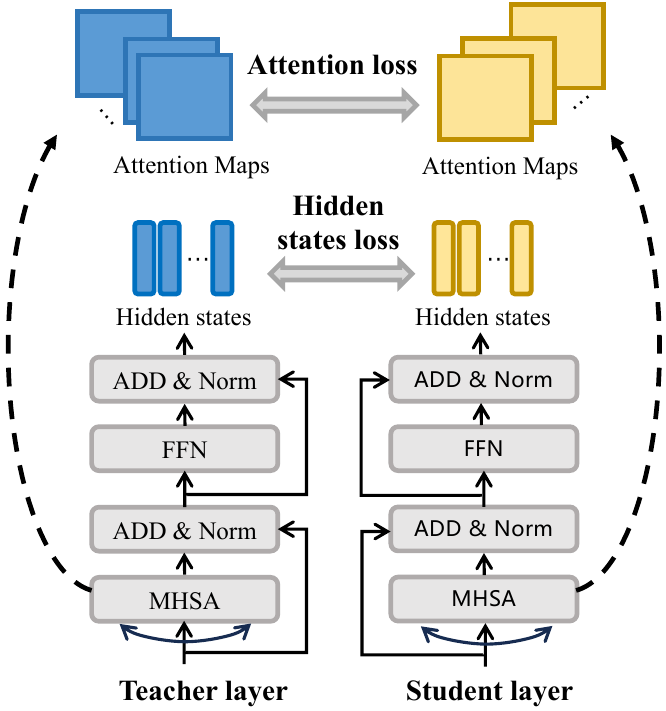}}
    \hfil
    \subfigure[Decoder knowledge distillation for hidden states and attention matrices]{
    \label{decoderkd}
    \includegraphics[width=1.9in]{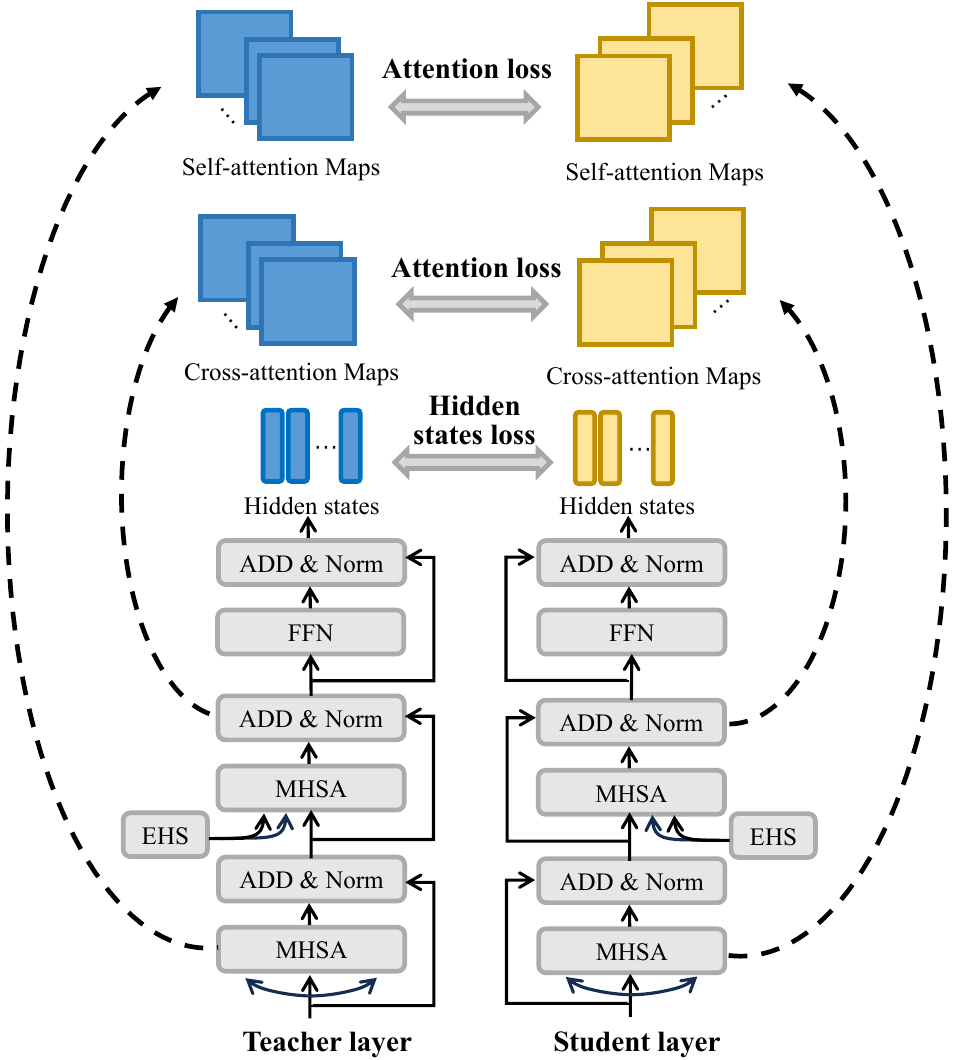}} 
    \hfil
  \caption{The details of Transformer layer for encoder and decoder feature-based knowledge distillation.}
  \label{encoderdecoderkd}
\end{figure}

\subsubsection{Feature-based KD} Deep neural networks learn multiple levels of feature representation from the input, and features of different intermediate layers represent varying degrees of knowledge \cite{Representationlearning}. Also, in contrast to the radio network, intermediate features learned from the audio network contain high-frequency bands and are devoid of noise due to the high quality of input.  
Therefore, in our design, we utilize the intermediate feature of the audio network to supervise the training of the radio network to conduct feature-based KD. 
In this process, the radio streaming Transformer mimics the high-quality feature representation of the audio streaming Transformer through the distillation loss that is specially designed for the Transformer, so that the missing high-frequency band can be complemented and the negative effect of noisy can be eliminated. 
As shown in Figure \ref{encoderdecoderkd}, we perform feature-based KD on attention matrices and hidden states of the encoder and decoder.

The attention weights in the Transformer have been proven to capture rich acoustic knowledge from input signals, which is essential for mmWave-based ASR \cite{attentionbert}. Therefore, we first use the attention-based KD to encourage the transfer of high-quality acoustic knowledge from the audio network to the radio network. Specifically, the knowledge transfer is achieved by the radio network mimicking the attention matrices of multi-head attention in the audio network. To facilitate the imitation of attention matrices, we design the loss function to penalize the distance between the attention matrices of the radio and audio streaming Transformers during teacher-student training. Consequently, the radio network can learn the acoustic feature with the high-frequency band and without noise, from the audio network with high-quality input. Considering Transformer contains three different attentions, including encoder self-attention (ESA), decoder self-attention (DSA), and decoder cross-attention (DCA), we define the object of attention-based KD as:
\begin{equation}
    \mathcal{L}_{att}=\mathcal{L}_{ESA}+\mathcal{L}_{DSA}+\mathcal{L}_{DCA}.
\end{equation}
Here $\mathcal{L}$ is the similarity function to compute the attention metrics loss and we select the mean squared error (MSE) to measure the distance between student and teacher models:
\begin{equation}
\begin{aligned}
        \mathcal{L}_{ESA}=\sum^{l_{enc}}_{i=1}MSE(ESA_i^T, ESA_i^S),\;
        \mathcal{L}_{DSA}=\sum^{l_{dec}}_{i=1}MSE(DSA_i^T, DSA_i^S),\; 
        \mathcal{L}_{DCA}=\sum^{l_{dec}}_{i=1}MSE(DCA_i^T, DCA_i^S),\;
\end{aligned}
\end{equation}
where $ESA_i$, $DSA_i$ and $DCA_i$ represent the attention matrix corresponding to the $i$-th ESA, DSA and DCA layer of teacher or student, and $l_{enc}$ and $l_{dec}$ refers to the number of encoder and decoder layers. 

Hidden states, as the output of Transformer layers, represent the learned feature of each layer. Therefore, hidden states also contain abundant acoustic knowledge learned from the input signals. Then, we use the hidden states based KD to transfer the high-quality acoustic knowledge from the audio network to the radio network, and it is an extension of attention-based KD for feature knowledge transfer. Specifically, we design the loss function to minimize the difference between the hidden states of radio and audio networks within teacher-student training, enabling the radio network to mimic the hidden states of each layer in the audio network. Through this process, the radio network can further learn to complement the missing high-frequency band and extract clearn features from low-quality input signals under the guidance of the audio network with high-quality input. The objective is to enhance the similarity of hidden states between audio network and radio network, and it can be formulated as 
\begin{equation}
    \mathcal{L}_{hid}=\mathcal{L}_{EHS}+\mathcal{L}_{DHS}.
\end{equation}
To achieve this goal, we use the MSE to measure the difference of hidden states in the encoder and decoder layers:
\begin{equation}
    \mathcal{L}_{EHS}=\sum^{l_{enc}}_{i=1}MSE(EHS_i^T, EHS_i^S), \; 
    \mathcal{L}_{EHS}=\sum^{l_{dec}}_{i=1}MSE(DHS_i^T, DHS_i^S),
\end{equation}
where the matrices $EHS_i$ and $EHS_i$ refer to the hidden states corresponding to the $i$-th encoder and decoder layer of teacher and student.

\subsubsection{Respondse-based KD} In addition to matching the intermediate features, we also directly imitate the behavior of the network response of the last output layer (i.e., logits). 
The performance of the audio network is superior to that of the radio network, as input audio signals possess a high-frequency band and are devoid of noise compared with radio signals. 
Therefore, it makes sense for the radio network to mimic the final prediction of the audio network. In response-based KD, the radio network is encouraged to fit the class probability distribution of each predicted word of the audio network. To achieve this, we utilize the distillation loss of soft logits to minimize the divergence between the soft logits of audio and radio networks. Specifically, we use the KL-Divergence to measure the difference between the soft logits of audio and radio networks and the objective is defined as:
\begin{equation}
    \mathcal{L}_{logits}=KLD(z^T/t,z^S/t),
\end{equation}
where $z^T$ and $z^S$ are the logits predicted by the student and teacher, respectively, and $t$ is the temperature to control the importance of each label and it converts output logits to soft targets.

\subsubsection{Cross-modal KD Training}
After initialization with the proposed GI, we combine the loss of the tailored streaming Transformer and the proposed cross-modal KD loss to train the network. Therefore, the overall training object can be formulated as
\begin{equation}
    \label{loss}\mathcal{L}=\alpha_{hybrid}\mathcal{L}_{hybrid}+\alpha_{att}\mathcal{L}_{att}+\alpha_{hid}\mathcal{L}_{hid}+\alpha_{logits}\mathcal{L}_{logits},
\end{equation}
where $\alpha_{hybrid}$, $\alpha_{att}$, $\alpha_{hid}$ and $\alpha_{logits}$ are hyper-parameters to balance loss terms. Once our tailored streaming Transformer is trained through the proposed cross-modal KD in the teacher-student paradigm, it only uses mmWave signals for inference. 

\section{Implementation} \label{Implementation}
\subsection{Experimental Setup}
We implement our system Radio2Text on a COTS and portable mmWave radar TIAWR1642BOOST \cite{awr1642} and an online data capture board DCA1000EVM \cite{dca1000evm}. The mmWave radar is modulated in FMCW with a frequency range of 77 GHz to 81 GHz to measure distance and angle. The streaming recognition is implemented by using the UDP protocol \cite{mmMesh}. The radar consists of two transmit (TX) antennas and four receive (RX) antennas. We use one TX antenna for transmission and four RX antennas for reception. In our configurations, the frame periodicity is set to 50 ms with 150 frames, and each frame contains 255 chirps with a duration of 180 us and a sampling point of 256. Therefore, after location with range FFT, the sample rate of mmWave signals (chirp sample rate) is 5.1 KHz. This sample rate is sufficient for the frequency range of intelligible human speech according to the Nyquist rate, i.e., $f_{mmWave}>2\times2.5 KHz$. Actually, as discussed in Section \ref{mmwavesignalsanalysis}, the main challenge is the upper frequency that mmWave radar can perceive is below 1.5 KHz. The configuration enables the radar to achieve a range resolution of 3.75 cm. Moreover, the mmWave radar has an azimuth FoV of 120$^{\circ}$ with a resolution of 15$^{\circ}$, which can encompass our experimental scenarios. 
With regard to the audio source, we use a loudspeaker that is commonly used in daily life to play the speech. After calibration, the length error between the original speech signals and mmWave signals is less than 10 ms.

\begin{figure}[!t]
  \centering
  \includegraphics[width=5.6in]{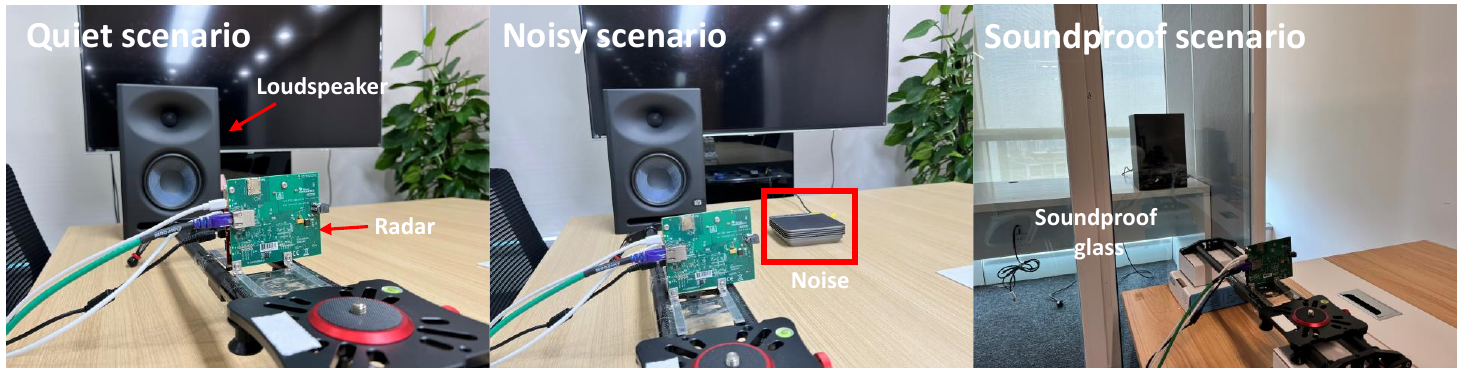}
  \caption{The experimental scenarios.}
  \label{scenarios}
\end{figure}

\subsection{Data Collection and Preparation} \label{collection}
We used the corpus of LJSpeech \cite{ljspeech} as the speech source for the loudspeaker. LJSpeech contains 13,100 speech clips with a mean duration of 6.57 seconds, and the total length is around 24 hours. Moreover, the corresponding transcriptions have 13,821 distinct words, and the mean words per clip are 17.23. To facilitate the training, words in the transcription are tokenized into subword units (aka character) using the SentencePiece toolkit \cite{sentencepiece}, and the character vocabulary size is set to 200. With the combination of characters, we can recover all the words in the transcription. According to the split of the widely used ASR toolkit ESPnet \cite{espnet}, we use the first 12,600 samples as the training set and the remaining as the test set. This division takes into account the fact that each speech clip in the test set contains not just a single word but multiple distinct words.

As shown in Figure \ref{scenarios}, we consider three scenarios, including the quiet scenario, the noisy scenario and the soundproof scenario. For the quiet scenario, the loudspeaker played the speech clips at a volume of 80 dB SPL, and the mmWave radar was placed in front of the loudspeaker at two distances of 50 cm and 150 cm to collect the corresponding mmWave signals. We selected these two distances as they are appropriate for the two respective applications of conference speech transcription and eavesdropping. Then, we included two more complex scenarios, namely noisy and soundproof scenarios. For the noisy scenario, an additional loudspeaker was placed around the radar to play noise at 60 dB SPL, and the distance between the radar and the original loudspeaker was 50 cm. In the soundproof scenario, the mmWave radar and loudspeaker were separated by a soundproof glass and the distance between them was 50 cm.
Experiments with different settings, such as varying distances between the mmWave radar and the loudspeaker, angles with respect to radar, loudspeaker volumes, and placements, were conducted in the quiet scenario. In each experiment, only one factor was changed while the other factors remained fixed.

The input pair of mmWave signals and speech signals are both resampled into a sample rate of 16 KHz. After preprocessing, mmWave signals are transformed into the 80-dimensional Mel-Spectrogram with a stride of 10 ms, and speech signals are transformed into the same output format. Moreover, to improve the network performance at different distances and angles, we propose a data augmentation method that linearly combines the Mel-Spectrogram $M_{radio}$ of mmWave signals with the Mel-Spectrogram $M_{bg}$ of background signals collected at different distances and angles, expressed as $M_{aug}=\beta_{radio}M_{radio}+\beta_{bg}M_{bg}$, where $\beta_{radio}$ and $\beta_{bg}$ are the coefficient and $\beta_{radio}+\beta_{bg}=1$. We conduct preliminary experiments to determine the coefficient value of $\beta_{radio}$. As the experimental value, $\beta_{radio}$ is randomly selected from the range [0.2, 0.6] due to the better performance.

\subsection{Network Setting and Training}

\subsubsection{Training Prerequisite}
\label{trainingprerequisite}
To enable the training of our model, we need to obtain the alignments for generating trigger events, the trained weight of the audio non-streaming Transformer teacher for initialization, and the trained audio streaming Transformer teacher for KD guidance. Therefore, we first train a CTC model, a non-streaming Transformer jointly trained with CTC and attention-based decoder, with mmWave signals as input and ground truth transcriptions as supervision. Subsequently, the well-trained CTC model uses Viterbi learning to compute the CTC path with the highest probability for input mmWave signals, providing alignment for further converting trigger events. 
Then, the audio non-streaming Transformer is trained as described above but with speech signals as input. It is served as the initialization of our tailored streaming Transformer. Also, the audio non-streaming Transformer provides the trigger events to facilitate training of the following audio streaming Transformer.
Next, we use speech signals as input and transcriptions as supervision to train the audio streaming Transformer, which serves as the teacher in the proposed cross-modal KD.

\subsubsection{Network Setting}
The audio streaming Transformer and the radio streaming Transformer share the same architecture, namely, the proposed tailored streaming Transformer in Section \ref{streamingtransformer}. It has 12 encoder layers and 6 decoder layers, with a head number of 8 and a self-attention dimension of 512 (please refer to \cite{transformer} for more details). In the encoder, the VGG-style CNN module has four identical blocks, each composed of a $3\times3$ convolution layer followed by a layer normalization and a ReLU activation. Also, a max pooling layer with a stride size of $2\times2$ is inserted after every two blocks. The chunk size in the encoder is 32, which is the trade-off between latency and accuracy. In the decoder, the CNN embedding layer has a stack of three blocks that is the same as the encoder.

The CTC model (radio non-streaming Transformer) and the network providing trained weights (audio non-streaming Transformer) described in Section \ref{trainingprerequisite} have the same architecture as our tailored streaming Transformer but without chunk-wise setting and triggered attention mechanism. 

\subsubsection{Training Setting}
The streaming Transformer is trained using the AdamW optimizer with a learning rate of 0.007 for 200 epochs. The weights assigned to different losses in Eq. \ref{hybrid} and Eq. \ref{loss} are set to $\alpha$=0.3, $\alpha_{hybrid}$=1, $\alpha_{att}$=3, $\alpha_{hid}$=1 and $\alpha_{logits}$=1. Regarding the non-streaming Transformer, it is trained using the Adam optimizer for 120 epochs, and the learning rate is set to 0.001. The weight in Eq. \ref{hybrid} is 0.3. We use PyTorch to implement all networks and train them with NVIDIA RTX 3090.

\section{Experiments Evaluation}
In this section, we conduct comprehensive experiments to evaluate the effectiveness and robustness of the proposed Radio2Text. We introduce evaluation metrics and baselines, and analyze the overall performance of our Radio2Text and compared methods. Then, we conduct the ablation study to verify each component. Finally, we evaluate our system under different settings.

\subsection{Evaluation Metrics} \label{metrics}
We use three quantitative metrics to evaluate the performance of our Radio2Text system in streaming ASR. 

\textbf{Word Error Rate (WER).} WER is a commonly used metric in ASR to evaluate recognition accuracy. It measures the difference between the transcription recognized by the system and the reference transcription. WER is calculated by dividing the number of word errors (i.e., insertions, deletions, and substitutions) by the total number of words in the reference transcription. The calculation can be represented as $WER=\frac{I_w+D_w+S_w}{N_w}$, where $I_w$ is the number of insertion errors (words in the recognized transcription that were not present in the reference transcription), $D_w$ is the number of deletion errors (words in the reference transcription that were not present in the recognized transcription), $S_w$ is the number of substitution errors (words in the reference transcription that were replaced by incorrect words in the recognized transcription), and $N_w$ is the total number of words in the reference transcription. A lower WER value indicates the better recognition performance of the ASR system. The WER can be greater than 1 when the number of errors in the recognized transcription is more than the number of words in the reference transcription.

\textbf{Character Error Rate (CER).} CER is another metric for ASR accuracy evaluation. Similar to WER, CER measures errors at the character level, and the calculation process can be expressed as $CER=\frac{I_c+D_c+S_c}{N_c}$, where $I_c$, $D_c$, $S_c$ and $N_c$ represent the number of insertion errors, deletion errors, substitution errors and characters in the reference transcription.

\textbf{Latency.} Latency is a critical metric to evaluate the real-time ability of a streaming ASR system. We use the frame-level latency to calculate the latency, which is the same as in \cite{tastreamingtransformer}. It is defined as the number of future frames that are consumed for each frame. Due to the usage of the chunk-wise setting, the latency of each frame is in a range between the latency of the start frame and the end frame. Therefore, we average the latency of each frame in a chunk to represent the latency metric.

\subsection{Baselines}
To demonstrate the effectiveness of our Radio2Text, we use the following baselines to compare with our system. First, we select the microphone as the comparator because audio signals are the most used modality for ASR. We utilize the audio-based streaming Transformer for comparison in the streaming setting, akin to the setting of Radio2Text. 
Then, mmWave-based ASR systems are selected for comparison, but they are all non-streaming methods.
mmWave-based non-streaming methods can be divided into two types: one directly utilizes mmWave signals directly for speech recognition, while the other utilizes the pre-recovered speech from mmWave signals for recognition. Although it is unfair to compare our system with those mmWave-based methods that solely work in non-streaming settings, this comparison can prove the superiority of our system on the other hand.

\textbf{Microphone+Streaming Transformer (Mic+S-Transformer).} The audio signals are collected from the microphone near the mmWave radar, and the trained audio streaming Transformer is leveraged for streaming speech recognition. Thus, Mic+S-Transformer is an audio-based streaming ASR system.

\textbf{Conformer RNN-T.} Conformer RNN-T without special design is a non-streaming network due to the usage of self-attention in Conformer \cite{conformer}. We follow the setting from AmbiEar \cite{ambiear} that directly uses Conformer RNN-T for speech recognition from mmWave signals in NLoS scenarios.  

\textbf{UNet+non-stremaing Transformer (UNet+NS-Transformer).} UNet uses mmWave signals to recover speech signals, which are then fed into the non-streaming Transformer for recognition. We reimplement the method described in \cite{radio2speech} and select the best results to recover the speech using the UNet-based method. Then, we employ the recovered speech to perform testing on the audio non-streaming Transformer. 

\textbf{cGAN+non-streaming Transformer (cGAN+NS-Transformer).} Similarly, the non-streaming Transformer is utilized to perform speech recognition from speech signals that are recovered using cGAN. We develop the same methods as in \cite{milliear} and use the best results of the cGAN-based method to recover speech. The recovered speech is used for testing on the audio non-streaming Transformer.

\begin{table}
  \caption{Examples of sentences decoded by Radio2Text and ground truth}
  \label{examplesentences}
  \begin{tabular}{c|c|c|c}
    \toprule
    ID &Mode &Groud Truth &Radio2Text \\
    \midrule
    \multirow{2}{*}{\thead{LJ050-0082}} &\thead{Word} &\thead[l]{the interest of the secret service goes beyond \\ information on individuals or groups threatening \\ to cause harm or embarrassment to the president} 
    &\thead[l]{the interest of the secret service goes beyond \\ information on individuals or groups threatening \\ to cause harm or embarrassment to the president} \\ 
    
    &\thead{Char} &\thead[l]{\_the \_in ter e st \_of \_the \_s e c re t \_s er v ic e \_g \\ o es \_be y on d \_in f or m ation \_on \_in d i v id u \\ al s \_or \_g r o u p s \_th re ate n ing \_to \_ca u se \\ \_h ar m \_or \_e m b ar ra s s ment \_to \_the \_president} 
    &\thead[l]{\_the \_in ter e st \_of \_the \_s e c re t \_s er v ic e \_g \\ o es \_be y on d \_in f or m ation \_on \_in d i v id u \\ al s \_or \_g r o u p s \_th re ate n ing \_to \_ca u se \\ \_h ar m \_or \_e m b ar ra s s ment \_to \_the \_president} \\
    \cmidrule{1-4}
    
    \multirow{2}{*}{\thead{LJ049-0019}} &\thead{Word} 
    &\thead[l]{the last presidential vehicle with any protection \\ against small arms fire left the white house in \\ nineteen fifty three} 
    &\thead[l]{the last presidential vehicle with any protection \\ against small arms fire left the white house in \\ nineteen fifty three} \\ 

    &\thead{Char} &\thead[l]{\_the \_l a st \_president i al \_v e h ic le \_with \_an y \\ \_pro t e c tion \_again st \_s m al l \_a r m s \_f i re \_le \\ f t \_the \_w h it e \_house \_in \_nineteen \_f if t y \_three} 
    &\thead[l]{\_the \_l a st \_president i al \_v e h ic le \_with \_an y \\ \_pro t e c tion \_again st \_s m al l \_a r m s \_f i re \_le \\ f t \_the \_w h it e \_house \_in \_nineteen \_f if t y \_three} \\
    \cmidrule{1-4}
    
    \multirow{2}{*}{\thead{LJ049-0128}} &\thead{Word} &\thead[l]{the fbi is the major *** DOMESTIC investigating \\ agency of the united states} &\thead[l]{the fbi is the major \textcolor{red}{THE} \textcolor{blue}{MESTIC}   investigating agency \\ of the united states} \\ 
    
    &\thead{Char} &\thead[l]{\_the \_f b i \_is \_the \_ma j or \_DO  M    e st ic \_in ve st i \\ g at ing \_a g en c y \_of \_the \_un it ed \_st ate s} &\thead[l]{\_the \_f b i \_is \_the \_ma j or \textcolor{blue}{\_THE \_M} e st ic \_in ve st i \\ g at ing \_a g en c y \_of \_the \_un it ed \_st ate s} \\
    \cmidrule{1-4}

    \multirow{2}{*}{\thead{LJ050-0136}} &\thead{Word} &\thead[l]{the committee WILL include representatives of the \\ president's office of SCIENCE  and technology \\ department of defense cia} &\thead[l]{the committee \textcolor{green}{****} include representatives of the \\ president's office of \textcolor{blue}{SIGNENCE} and technology \\ department of defense cia} \\ 
    
    &\thead{Char} &\thead[l]{\_the \_co m m it t e e \_W ILL \_in c l u d e \_re p re \\ s ent at ive s \_of \_the \_president ' s \_office \_of \_s \\ C i * * ence \_and \_t e ch n o l o g y \_de p ar t ment \\ \_of \_de f en se \_c i a} &\thead[l]{\_the \_co m m it t e e \textcolor{blue}{**** ***} \_in c l u d e \_re p re \\ s ent at ive s \_of \_the \_president ' s \_office \_of \_s \\ \textcolor{green}{*} i \textcolor{red}{G N} ence \_and \_t e ch n o l o g y \_de p ar t ment \\ \_of \_de f en se \_c i a} \\

    \bottomrule
\end{tabular}
\end{table}

\subsection{Overall Performance}
This section analyzes the effectiveness of the proposed Radio2Text in terms of streaming ASR. To further study the robustness of the system in complex scenarios, we evaluate its performance in noisy and soundproof scenarios.

\begin{table}[t]
  \caption{Quantitative evaluation results of Radio2Text at distances of 50 cm and 150 cm}
  \label{quantitativeevaluation}
  \begin{tabular}{ccccc}
    \toprule
    Method &Collection Distance (cm) &CER (\%) &WER (\%) &Latency (ms) \\
    \midrule
    \multirow{2}{*}{Radio2Text} &50  &5.7 &9.4 &\multirow{2}{*}{640} \\
        &150 &10.2 &15.4 \\
    \bottomrule
\end{tabular}
\end{table}

\subsubsection{Streaming Recognition Performance}
The streaming recognition performance is evaluated from qualitative and quantitative perspectives.
To give an intuitive awareness of the recognition performance, we provide some examples of ground truth and sentences decoded by Radio2Text, both in character and word levels, as shown in Table \ref{examplesentences}. The majority of the decoded sentences are accurate, and we especially select certain representative misrecognized sentences for reference. The recognized sentences are aligned with the ground truth for calculating errors. Capitalized words/characters represent a mismatch between the ground truth and recognized sentence, while ``***" indicates the absence of certain words/characters. Also, the colors \textcolor{blue}{blue}, \textcolor{green}{green}, and \textcolor{red}{red} correspond to the error types of substitution, deletion, and insertion, respectively, as explained in Section \ref{metrics}. It is obvious that the sentences containing many distinct words are correctly recognized, and the length of sentences does not impinge the recognition performance. Even for the error instances, misrecognized words/characters are very close to the ground truth (e.g., DOMESTIC $\to$ MESTIC, SCIENCE $\to$ SIGNENCE), and the incorrect insertion and deletion of characters/words are normally articles, which do not affect the understanding of sentences. The probable reason is that inarticulate pronunciations of utterances cause confusion for our system. The results indicate the ability of the proposed Radio2Text to accurately recognize the text with many distinct words beyond just simple digital or commands.

In quantitative evaluation, we use three metrics, namely WER, CER and Latency, to measure the performance of Radio2Text. We conduct our evaluations at two distances, i.e., 50 cm and 150 cm, which correspond to two potential applications of conference speech transcription and eavesdropping. The former allows the radar to put in a close distance, and the attack distance for the latter is normally set to that distance \cite{mmEavesdropper}. As shown in Table \ref{quantitativeevaluation}, our Radio2Text achieves a CER and WER of 5.7\% and 9.4\% at a distance of 50 cm, which basically meets the requirement of conference transcription. Although the error rate increases at a distance of 150 cm, our system still maintains a relatively low error rate, which demonstrates the feasibility of using mmWave-based streaming ASR for eavesdropping. For the latency, considering the down-sampling rate of VGG is 4, the input feature sampling rate is 10 ms and the chunk size is 32, the maximum latency is calculated as $32\times4\times10=1280 ms$. Thus, the latency of each frame in the encoder is in the range of [0, 1280], and the average latency is 640 ms, which is an acceptable latency in the audio domain \cite{scout}. Since mmWave signals collected from these two distances are all evaluated by the same network, they possess the same latency of 640 ms. 
These results validate the effectiveness of Radio2Text in utilizing mmWave signals to accurately recognize the text containing a vocabulary of over 1,3000 words in a streaming mode, which provides the possibility for conference speech transcription and eavesdropping.

\begin{figure}[!t]
  \centering
  \includegraphics[width=\linewidth]{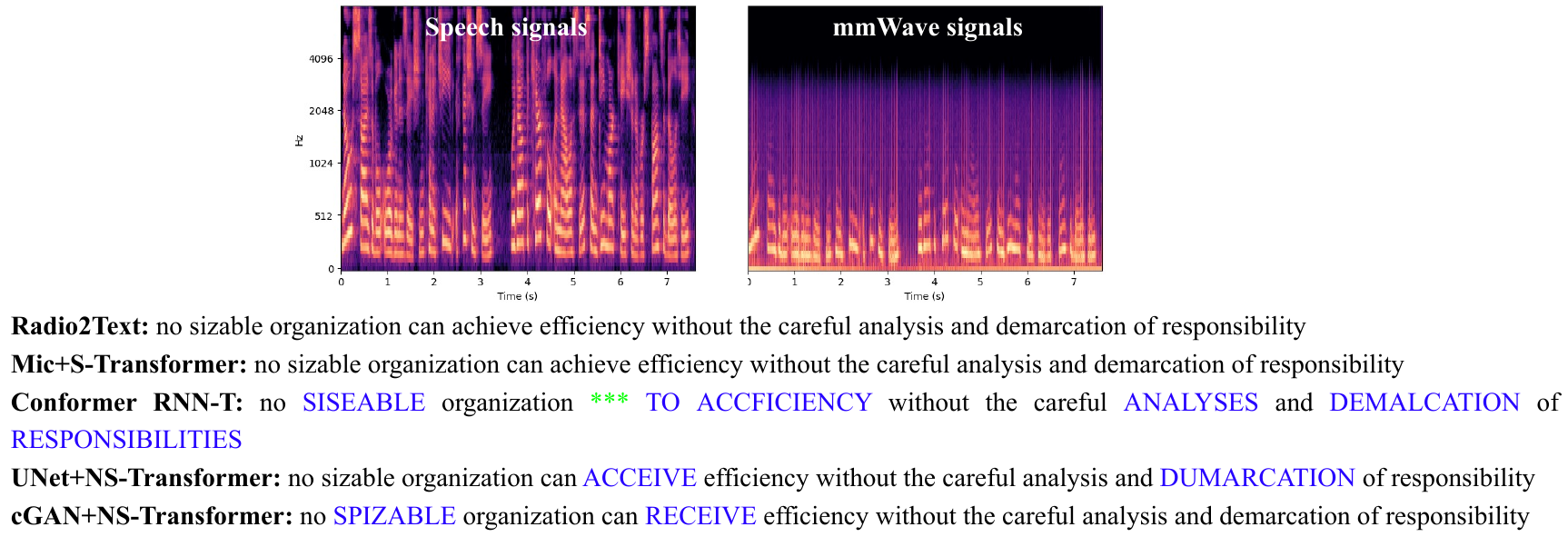}
  \caption{Mel-Spectrograms of speech and mmWave signals, and the corresponding recognition results of microphone-based streaming ASR method and mmWave-based non-streaming ASR methods. }
  \label{overallresults}
\end{figure}

\subsubsection{Comparison with Baselines}
To further demonstrate the effectiveness of Radio2Text in recognizing low quality mmWave signals, we compare our Radio2Text with other methods, including microphone-based streaming ASR methods and other mmWave-based non-streaming ASR methods. The experiments in this section are conducted on mmWave signals at 50 cm. We first visualize the input signals and corresponding recognition results for an intuitive presentation. As shown in Figure \ref{overallresults}, although the quality of mmWave signals is inferior to that of speech signals in terms of Mel-Spectrogram, it is gratifying that Radio2Text can accurately recognize text from low quality mmWave signals just like recognizing audio from a microphone. However, other mmWave-based methods yield certain erroneous words, which indicates that Radio2Text has a superior ability to recognize text from low quality mmWave signals.

Moreover, the quantitative comparison results are shown in Table \ref{comparison}. Only the microphone-based method, \emph{Mic+S-Transformer}, is compared for streaming ASR because other existing mmWave-based methods do not support streaming ASR. The performance of our Radio2Text is on par with that of the microphone-based method in terms of error rate. 
We also include mmWave-based non-streaming ASR methods for comparison, including \emph{Conformer RNN-T}, \emph{UNet+NS-Transformer} and \emph{cGAN+NS-Transformer}. Indeed, it is unfair to compare Radio2Text with those non-streaming methods, as the latter have access to the entire future input for prediction. We expect this comparison can prove the superiority of our system on the other hand.
Although it is intuitive that using the recovered speech for recognition will be more accurate than directly using mmWave signals for recognition, the results show \emph{UNet+NS-Transformer} and \emph{cGAN+NS-Transformer} are actually inferior to ours by at least 9.8\% and 12.5\% for CER and WER. This is because the recognition performance of such methods highly relies on the quality of recovered speech, and certain speech-related information is inevitably lost during the recovery process. As systems that both use mmWave signals directly for recognition, Radio2Text outperforms \emph{Conformer RNN-T} by 16.7\% and 25.1\% for CER and WER, respectively. This indicates that directly using mmWave signals for recognition is difficult without the proposed component of GI and cross-modal KD.

These results manifest that Radio2Text has a comparable ability with the microphone in streaming recognition, and demonstrate the superiority of Radio2Text over other mmWave-based non-streaming ASR methods. This indicates that Radio2Text has the ability to comprehensively explore the information in low quality mmWave signals, resulting in superior recognition results.

\begin{table} [t]
  \caption{Comparison results of Radio2Text and baselines}
  \label{comparison}
  \begin{tabular}{c|ccc}
    \toprule
    Mode &Method &CER (\%) &WER (\%) \\
    \midrule
    \multirow{2}{*}{Streaming} &\textbf{Radio2Text (ours)} &\textbf{5.7} &\textbf{9.4} \\
    &Mic + S-Transformer &4.1 &7.6 \\ 
    \cmidrule{1-4}

    \multirow{3}{*}{Non-Streaming} &Conformer RNN-T &22.3 &34.5 \\
    \cmidrule{2-4}
    &UNet + NS-Transformer &15.5 &21.9 \\ 
    &cGAN + NS-Transformer &16.3 &23.3 \\  
    
    \bottomrule
\end{tabular}
\end{table}

\begin{table}
  \caption{Evaluation in complex scenarios}
  \label{complexscenarios}
  \begin{tabular}{c|c|cc|cc}
    \toprule
    \multirow{2}{*}{Mode} &\multirow{2}{*}{Method} &\multicolumn{2}{c|}{Noisy Scenario} &\multicolumn{2}{c}{Soundproof Scenario} \\
    & &CER (\%) &WER (\%) &CER (\%) &WER (\%)\\
    \midrule

    \multirow{2}{*}{Streaming} &\textbf{Radio2Text (ours)} &\textbf{5.8} &\textbf{9.6} &\textbf{19.6} &\textbf{26.5} \\
    &Mic + S-Transformer &67.2 &72.5 &54.3 &65.8\\ 
    \cmidrule{1-6}

    \multirow{2}{*}{Non-Streaming} &Conformer RNN-T &22.1 &34.4 &43.2 &51.4 \\
    \cmidrule{2-6}
    &UNet + NS-Transformer &15.8 &22.0 &25.1 &33.2 \\
    &cGAN + NS-Transformer &16.1 &23.5 &28.3 &38.0 \\ 
    
    \bottomrule
\end{tabular}
\end{table}

\subsubsection{Complex Scenarios Analysis}
In practical application, working scenarios may not always be quiet, and it is common to encounter noise and soundproof materials in such scenarios. Therefore, we further evaluate the effectiveness and robustness of Radio2Text in complex scenarios, including two common scenarios: noisy and soundproof scenarios. In the noisy scenario, the audio that is not relevant to the target source is considered as noise and is played by another loudspeaker outside the detection range of mmWave radar. As for the soundproof scenario, a soundproof glass is placed between the mmWave radar and the loudspeaker. We use the data collected in noisy and soundproof scenarios for training and testing, as explained in Section \ref{collection}.

The quantitative results are shown in Table \ref{complexscenarios}. In noisy scenarios, the microphone-based streaming ASR method (i.e., \emph{Mic+S-Transformer}) almost fails, with a CER of 67.2\% and a WER  of 72.5\%, In contrast, mmWave-based methods still maintain a similar performance to that in the quiet scenario, and Radio2Text achieves the best performance in terms of CER and WER. The phenomena can be attributed to the sensitivity of microphones to noise, which has no effect on radio signals. In the soundproof scenario, despite a decrease in performance compared to quiet and noisy scenarios, Radio2Text still achieves the lowest error rate among all streaming and non-streaming ASR methods. This is because of the attenuation of radio signals caused by the soundproof glass. The microphone-based streaming ASR method still performs the worst, since the soundproof glass block part of speech signals. This clearly shows that our Radio2Text is robust to complex scenarios, such as noisy and soundproof scenarios, an ability that microphones do not possess. 

\begin{table}
  \caption{Ablation study results}
  \label{ablation}
  \begin{tabular}{c|cc}
    \toprule
    Method &CER (\%) &WER (\%)\\
    \midrule
    \textbf{Radio2Text (Ours)} &\textbf{5.7} &\textbf{9.4} \\
    \cmidrule{1-3}
    w/o GI \& Cross-modal KD &21.1 &32 \\
    w/o Tailored streaming Transformer &33.2 &45.1  \\
    w/o GI &9.1 &15.9 \\
    w/o Cross-modal KD &10.4 &16.8 \\
    \cmidrule{1-3}
    w/o Feature-based KD &8.2 &13.2 \\
    w/o Response-based KD &7.4 &12.1 \\
    \bottomrule
    
\end{tabular}
\end{table}

\subsection{Ablation Studies} \label{ablationstudy}
In this section, we perform some ablation studies to validate the effectiveness of each component in Radio2Text. All experiments were carried out on the testing set of LJSpeech collected at 50 cm in the quiet scenario. The results are shown in Table \ref{ablation}.

\textbf{w/o GI \& Cross-modal KD} represents only using the tailored streaming Transformer with mmWave signals as the input, without the utilization of GI and cross-modal KD. The results show that CER/WER significantly increase from 5.7\%/9.4\% to 21.2\%/32\%. This indicates the proposed GI and cross-model KD can effectively address the issues caused by low quality input mmWave signals and limitations of streaming ASR setting, improving streaming recognition performance.

\textbf{w/o Tailored streaming Transformer} means the tailored streaming Transformer is replaced by a typical encoder-decoder-based bidirectional LSTM. Considering the cross-modal KD is designed for the Transformer structure, we do not use it for training that LSTM. The error rate of CER and WER increases significantly without the tailored streaming Transformer, which demonstrates the powerful ability of the tailored streaming Transformer to extract feature representation. 

\textbf{w/o GI} refers to not using the proposed GI to initialize the network weights but using the random initialization. The results show that not using the GI results in 3.4\% and 6.5\% gain on CER and WER. This manifests that the feature knowledge of global context inherited from the non-streaming Transformer through the proposed GI is important for improving recognition performance.

\textbf{w/o Cross-modal KD} represents that the proposed cross-modal KD is removed. We find that the error rate of CER/WER increase from 5.7\%/9.4\% to 9.5\%/15.9\%. This indicates that the feature and response knowledge transferred from audio modality to radio modality can facilitate the training, improving the recognition accuracy.

\textbf{w/o Feature-based KD} means that we remove the feature-based KD from cross-modal KD. The results show that feature-based KD provides 2.5\% and 3.8\% gain on CER and WER, which demonstrates the effectiveness of the intermediate feature provided by the teacher network.

\textbf{w/o Response-based KD} represents that we remove the response-based KD from cross-modal KD. It can be seen the CER and WER are increased to 7.4\% and 12.1\%, which indicates that the response guidance from the teacher network is valuable for student prediction, allowing the student to make accurate predictions.

\subsection{Impact of Different Factors}
In this section, we analyze the robustness of Radio2Text under different distances between radar and loudspeaker, radar orientations, volume levels, and environmental changes.

\begin{figure}[t]
  \centering
    \subfigure[Impact of distance between radar and loudspeaker]{
    \label{impactdistance}
    \includegraphics[width=1.5in]{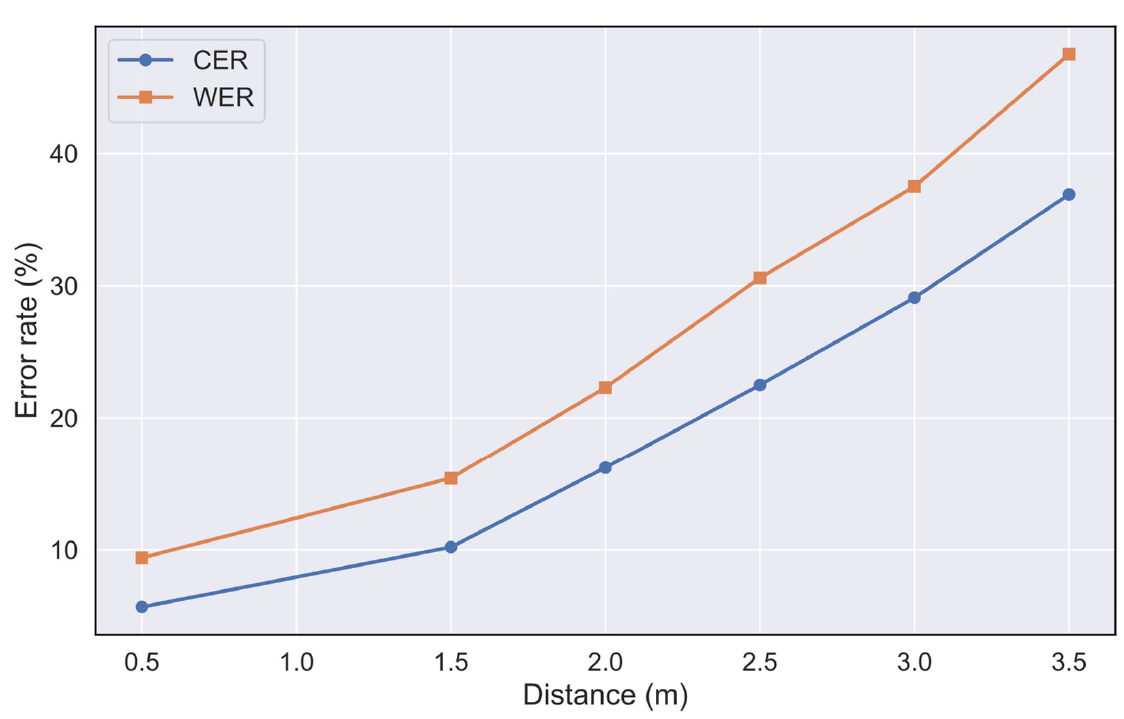}} 
    \hfil
    \subfigure[Impact of radar orientation]{
    \label{impactorientation}
    \includegraphics[width=1.5in]{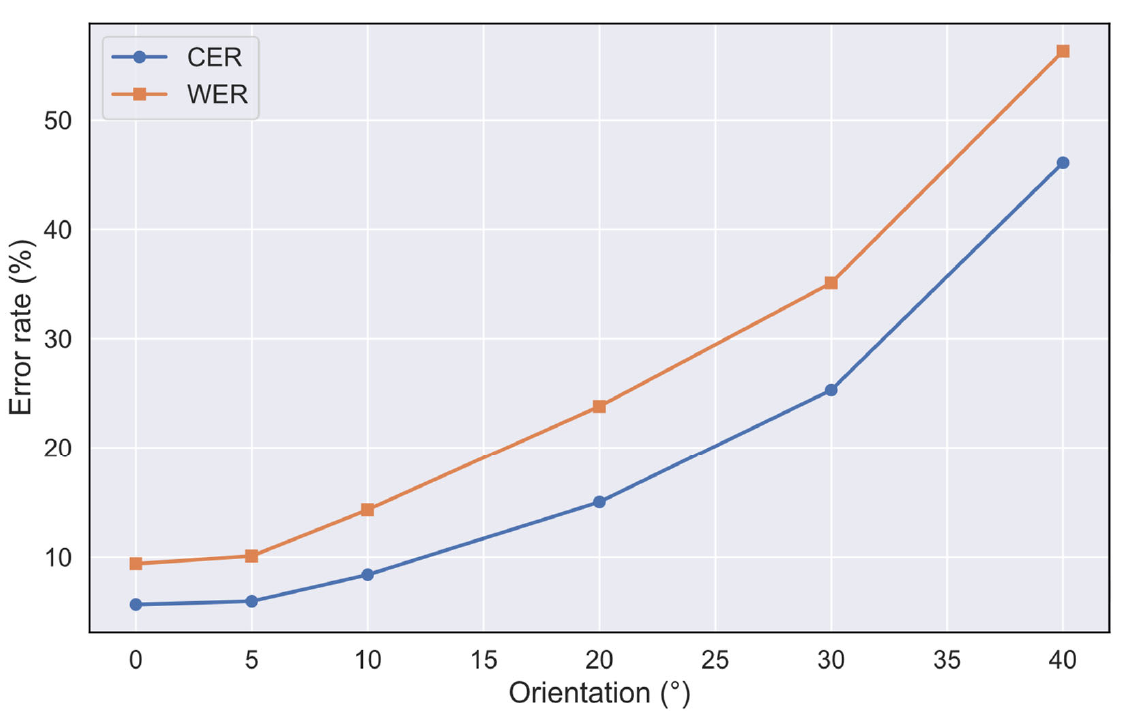}} 
    \hfil
    \subfigure[Impact of volume level]{
    \label{impactvolume}
    \includegraphics[width=1.5in]{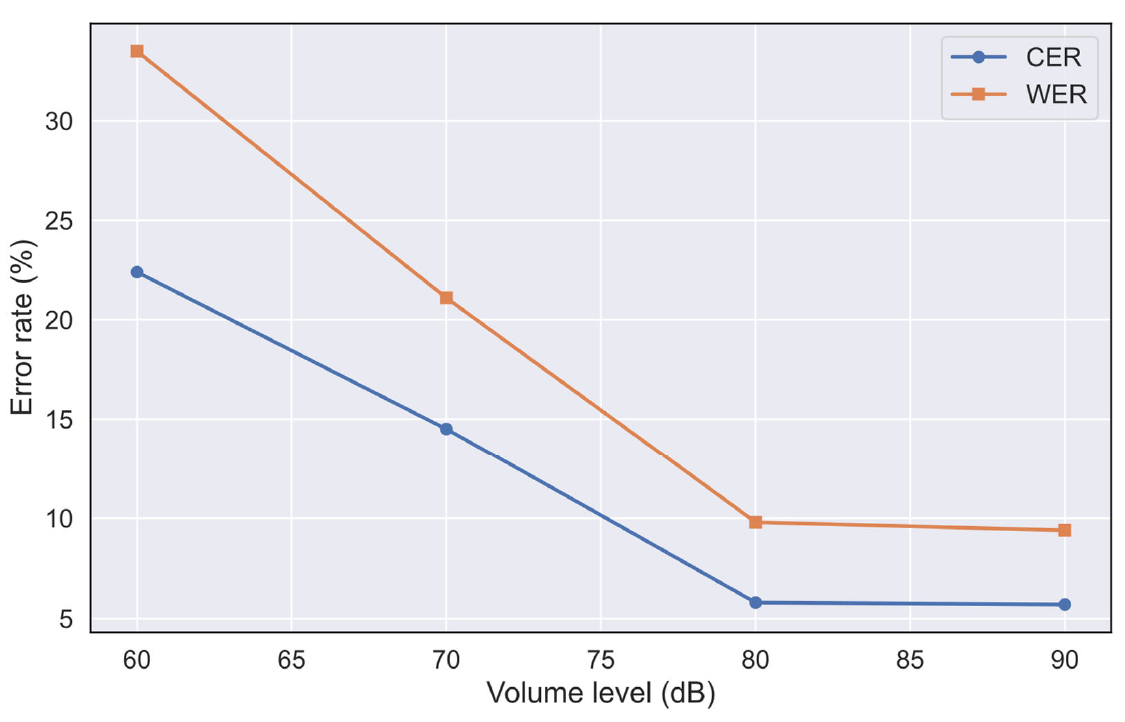}} 
    \hfil
    \subfigure[Impact of environmental change]{
    \label{impactplacement}
    \includegraphics[width=1.4in]{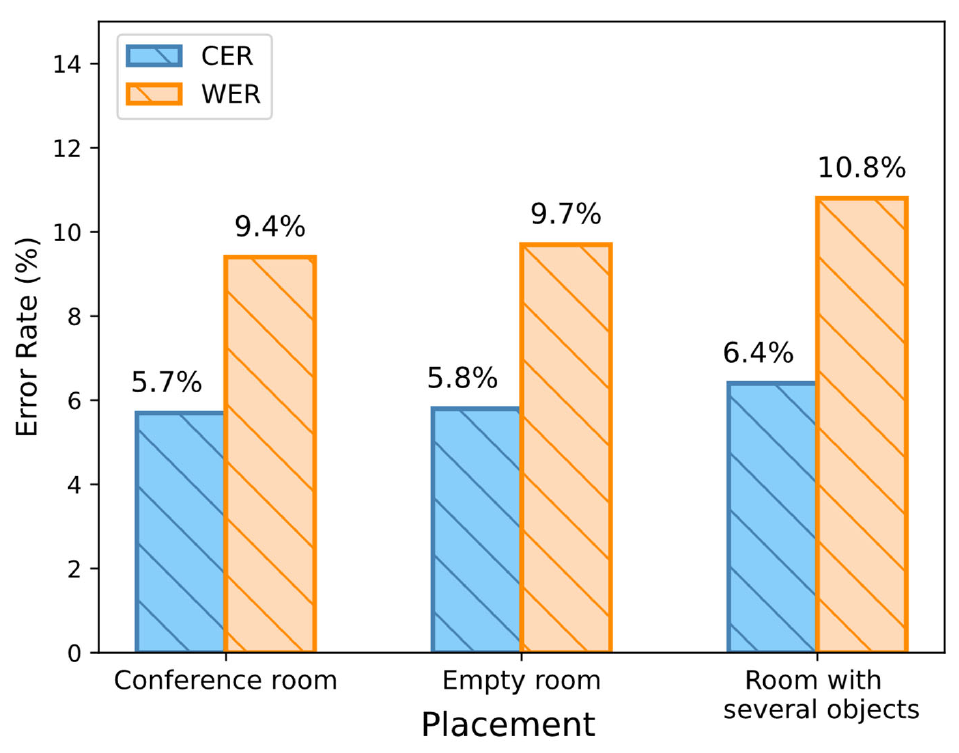}}
    \hfil
  \caption{Impact of different factors.}
  \label{otherfactors}
\end{figure}

\subsubsection{Distance between Radar and Loudspeaker}
In this experiment, we evaluate the performance of Radio2Text at different distances between mmWave radar and loudspeaker. We keep the loudspeaker in a fixed position and move the radar so that the distance between them varies from 0.5 m to 3.5 m. The experimental results are shown in Figure \ref{impactdistance}. We can find that the CER and WER of Radio2Text increase as the distance between radar and loudspeaker increases. This is because the vibration amplitude of the loudspeaker is inherently weak and the SNR increases as the distance increases. Therefore, as the propagation of mmWave signals increases, the attenuation of speech-related mmWave signals becomes increasingly severe, making them susceptible to interference from noise. This is more challenging for streaming recognition with a large vocabulary. In general, our Radio2Text can maintain a relatively low CER, below 20\%, within a distance of 2.0 m.

\subsubsection{Radar Orientation}
In real scenarios, the radar may not be directly facing the loudspeaker. Therefore, we also investigate the performance of Radio2Text at different angles of radar towards the loudspeaker. In this experiment, we change the radar orientation from 0 degrees to 40 degrees at a distance of 0.5 m without changing the position of the loudspeaker. As shown in Figure \ref{impactorientation}, the error rate exhibits a relatively smooth increase as the orientation expands, as long as the orientation remains within 30 degrees. But, the error rate increases drastically in the orientation of 40 degrees. The reason is that the vibration area that is visible for radar becomes increasingly smaller as the orientation angle increases. Thus, the reflected mmWave signals contain progressively lesser speech-related information, which results in an increase in recognition errors. Generally, our Radio2Text is able to keep the CER below 20\% in the orientation of less than 30 degrees.

\subsubsection{Volume Level}
In addition to distance and orientation, the volume level is another factor that directly affects the sensing quality. To evaluate the performance of Radio2Text in different volume levels, we change the volume of the loudspeaker at a fixed distance of 0.5 m and orientation of 0 degrees. The volume level of the loudspeaker is varied from 60 dB to 90 dB, with an environment volume level of 30 dB. Figure \ref{impactvolume} shows the error rate under different volume levels. We can find that CER and WER increase with the volume level decreases. This is because as the volume decreases, the loudspeaker's vibration amplitude also weakens. In addition, the displacement resolution that mmWave radar can perceive is limited. Therefore, the amount of speech-related information that the mmWave radar can perceive reduces, resulting in a decline in performance.

\subsubsection{Environmental Change}
The presence of background objects in the room can also reflect mmWave signals, which may generate static clutter and multipath signals. To investigate the impact of environmental changes on Radio2Text, we conduct experiments in three different scenarios, including a conference room, an empty room and a room with several objects. The conference room contains some chairs and tables, while the room with several objects contains a table and several objects on the table, such as books and bottles. The results of the error rate are shown in Figure \ref{impactplacement}. It can be seen that the error rate is stable across these three scenarios, and there is a slight increase in the error rate in the last scenario. This is because our Radio2Text uses some signal processing to reduce the effect of static clutter, and the proposed cross-modal KD can alleviate the negative effect of low quality input signal to some extent.

\section{Discussion and Future Work}
\subsection{Influence of Vocabulary Size}
The influence of vocabulary size on ASR models is twofold. On the one hand, a larger vocabulary size allows for wider and more practical usage. This is because the ASR model with a larger vocabulary size is capable of understanding a wider range of words so that the model can respond more accurately to users' requests and adapt to versatile users. On the other hand, a large vocabulary size increases the potential for matches, raising the level of uncertainty in recognition. This necessitates the model to possess a powerful feature representation ability to deal with the increased uncertainty. To train such models, a larger corpora dataset for speech recognition is also required. In the audio domain, extensive large speech corpora with thousands of hours are accessible \cite{1000language,librispeech}, and with the help of the large dataset, powerful speech representation models for self-supervised learning are trained to promote ASR \cite{wav2vec2,hubert}. However, there exists no large dataset of speech-related mmWave signals due to the difficulty of data collection, which limits the further scale of mmWave-based ASR models to a large vocabulary. 
Although the proposed GI and cross-modal KD facilitate the streaming Transformer to recognize over 1,3000 words with limited training data (relative to large audio datasets), the assistance they provide is finite. Therefore, further enlarging the vocabulary size without sufficient training data renders the powerful network susceptible to overfitting, which increases the ambiguity in recognition for our streaming Transformer and diminishes the recognition performance. As advancements in millimeter wave-based sensing, we believe a larger dataset of speech-related mmWave signals will emerge in the feature. In that scenario, the mmWave-based ASR model could unleash its full potential for feature representation, given sufficient training data, thereby expanding the recognizable vocabulary size. Establishing a large dataset of mmWave signals and training a feature representation model based on the large dataset through self-supervised learning would be a part of our future work.

\subsection{Multi-target Scenarios}
In certain situations, it is common for multiple sound sources to produce sound simultaneously, named cocktail party effect \cite{cocktail}. When there is more than one sound source within the detection scope of mmWave radar, the vibration of each target will be captured by the radar. Specifically, the reflected signals of radar are the mixture of the vibration signals reflected from different targets. To achieve multi-speaker ASR, it is natural to use a two-step procedure in which the mixed speech-related mmWave signals are first separated, and the mmWave-based ASR model is then performed on each separated signal. When two targets are close, it is difficult to separate the individual signal from the mixed reflection signal by directly distinguishing the signals from different range bins. In this context, some blind source separation methods \cite{blindsourceseparation} and speech separation neural networks \cite{dlseparation} can be used for signal separation. Then, our Radio2Text can still be used for the recognition of separated signals. An alternative approach for multi-speaker ASR is to design a neural network to perform multi-speaker speech recognition based on mmWave signals without an explicit separation module. Multi-speaker ASR would be explored in our future work.

\subsection{Ethical Consideration for Potential Applications}
The proposed Radio2Text has the potential to be applied for eavesdropping and transcription. For example, when attending an online meeting or communicating with a voice assistant, the mmWave radar can be placed in front of the loudspeaker to transcribe the speech for recording, whether the scenario is quiet or noisy, and it can even be placed behind soundproof materials for eavesdropping. For the application for conference speech transcription, mmWave-based ASR can effectively improve recognition accuracy in a noisy scenario where microphones almost fail. This approach provides a novel possibility for transcription and offers a means of enhancing user experience in noisy scenarios. Also, mmWave-based ASR demonstrates the potential for eavesdropping, even in noisy and soundproof scenarios. It is worth noting that the usage of mmWave-based ASR technology may raise concerns about privacy and security. People should be vigilant even in a soundproof room or where there is no microphone. Considering these risks, there are some solutions to defend against the ``mmWave attack". For example, the usage of RF shielding (e.g., metal block) is a possible countermeasure as RF signals cannot penetrate them.

\section{Related Work}
In this section, we introduce the related work from three aspects: sensor- and vision-based audio sensing, radio-based audio sensing and cross-modal knowledge distillation.

\textbf{Sensor- and vision-based audio sensing.}
Numerous types of sensors have been proven to recover audio information by detecting sound-included vibration displacement, e.g., Inertial Measurement Units (IMU) \cite{gyrophone, accear, inertiear}, vibration motors \cite{listening}, wireless signals\cite{wei2015acoustic} and even magnetic hard drives \cite{hard}, etc. Michalevsky et al. \cite{gyrophone} and Hu et al. \cite{accear} develop sound recovery systems based on gyroscopes and accelerometers respectively. Additionally, InertiEAR \cite{inertiear} combines both accelerometers and gyroscopes together and exploits coherence between responses of inertial sensors and hardware diversity using a mathematical model in order to make automatic audio sensing more precise. Roy and Choudhury \cite{listening} demonstrate a vibration motor-based system to capture the sound around the devices. Kwong et al.\cite{hard} eavesdrop by hard disk drives. However, these sensors still suffer from the limitation of low sampling frequency. Besides, vision based audio sensing also developed a lot. Researchers use high-speed cameras to capture the vibration of tiny objects in the video \cite{visualmicrophone}. Also, \cite{spying} implements a method to recover sound via the Lidar equipped on the robot. Unfortunately, the performance of the vision based system is highly dependent on illumination conditions.

\textbf{Radio-based audio sensing.}
RF signals can penetrate through obstacles and diffract around them, facilitating sensing even in low illumination conditions, which can make up for the lack of the above-mentioned sensors. Researchers have shown that it is possible to reconstruct audio using RF signals by detecting vibrations \cite{mmEavesdropper, mmmic, mmspy, mmeve, mmphone, milliear, mmecho, radio2speech, ambiear, radioses, mmvib,wavesdropper, waveear}. Wei et al. \cite{wei2015acoustic} implement an audio sensing system by WiFi antenna array as the pioneer for RF-based speech recovery. After that, \cite{mmvib} and \cite{mmecho} proposes the mmWave-based acoustic sensing system without the aid of machine learning or prior knowledge. In research of \cite{mmphone}, an RF signal-based system is proposed to decode speech from piezoelectric film and enhance the quality of speech. mmEve \cite{mmeve} is an end-to-end eavesdropping system that can be controlled remotely and is motion-resilient. Also, \cite{radioses} implements a joint audio-radio speech enhancement and separation system using mmWave sensing. \cite{radio2speech} proposes the Radio UNet to recover high quality speech from the time-frequency domain in different scenarios. Feng et al. \cite{mmEavesdropper} propose a mmWave-based solution to eavesdrop via micro-vibration signal from three aspects. mmMIC \cite{mmmic} performs well in speech recognition by detecting the tiny vibrations of lips with the random motion of the body. Additionally, Basak et al. \cite{mmspy} eavesdrop phone calls by detecting minute vibrations produced by the earpiece device. Zhang et al. \cite{ambiear} propose a system to eliminate limitations on human position and posture while ensuring high quality of speech in non-line-of-sight scenarios. 

Some of the above-mentioned works conduct a speech recognition experiment to test their performance. However, they mainly focus on the simple task of word/command classification, and the number of recognizable words is limited. As shown in Table \ref{distinctwords}, they primarily recognize command-like words, such as digits, letters, and specific commands, making it more akin to a classification task, instead of the complex task of speech recognition that recognizes the sentence containing many distinct words. Most importantly, they can not perform streaming speech recognition. If mmWave-based speech recognition systems are to be applied in practice, the delay and the size of the recognizable vocabulary are critical factors. 

\begin{table}[t]
  \caption{Comparison with previous works}
  \label{distinctwords}
  \begin{tabular}{c|c|c|c}
    \toprule
    Works & Dataset &{\# of Distinct Words} &\makecell{Support \\ Streaming ASR}\\
    \midrule
    mmEve\cite{mmeve} & {Harvard Sentences (part of) \cite{harvard}} & \textit{approx.} 460 &\ding{55} \\
    mmPhone\cite{mmphone} & \makecell{AudioMNIST \cite{audiomnist} $\&$  \\ Harvard Sentences (part of) \cite{harvard}} & \textit{approx.} 470 &\ding{55} \\
    WaveEar\cite{waveear} & standardised reading materials &814 &\ding{55} \\
    mmSpy\cite{mmspy} & \makecell{AudioMNIST \cite{audiomnist} $\&$ \\ Speech Commands \cite{commands}} & 40 &\ding{55} \\ 
    Wavesdropper\cite{wavesdropper} & self-designed dataset & 57 &\ding{55} \\
    mmMIC\cite{mmmic} & {phonetic symbols} & 48  &\ding{55} \\
    mmEavesdropper\cite{mmEavesdropper} & {digits and letters} & 36  &\ding{55}  \\
    AmbiEar\cite{ambiear} & Speech Commands \cite{commands} & 30 &\ding{55}\\
    \cmidrule{1-4}
    \textbf{Radio2Text (Ours)} & \textbf{LJSpeech} \cite{ljspeech} & \textbf{13,100} &\ding{51} \\
    \bottomrule
\end{tabular}
\end{table}

\textbf{Cross-modal knowledge distillation.}
KD is primarily designed for model compression in the initial concept \cite{kd}. Its general idea is to transfer knowledge from a teacher network to a student network.
As a result of its versatility, KD has been widely applied in various tasks, such as semantic segmentation \cite{semanticseg1,semanticseg2}, object detection \cite{objectdetection1, objectdetection2}, and pre-trained language model compression \cite{bertcompress1, bertcompress2}. Different categories of knowledge are investigated for distillation, and the widely used strategy is to match the soft targets of the output layer \cite{kd}, intermediate features \cite{FitNets, AttentionTransfer} and the relation of specific layers. Considering that real-world data is associated with different modalities, KD has also been extended to transfer knowledge in multi-modal data \cite{multimodalitykdanalysis}. In \cite{asrallyouneed}, a visual speech recognition model is trained by distilling the knowledge from a pre-trained ASR model with paired input audio. Similarly, through a teacher-student training framework, a well-established visual recognition model transfers knowledge into SoundNet using unlabeled video (sound + vision) \cite{soundnet}. Moreover, researchers leverage the mid-level features learned by CNNs trained on images to supervise the training of a CNN on depth images \cite{depthkd}. In addition, RF signals can also serve as recipients of the knowledge transferred from visual signals. For example, in \cite{humanpose}, the radio-based network uses the pose information generated from visual signals as the supervision to guide the training. Also, RF-Diary \cite{rfdiary} aligns its features to the same space of features extracted from a video-captioning model to transfer the visual knowledge of event captioning to RF signals. 

Our work falls in the field of cross-modal KD, which transfers the knowledge learned in one dominant modality to another. Unlike previous RF-based systems that leverage the final output of the other modality to supervise the training, we transfer not only the knowledge of the output layer but also the knowledge of intermediate layers that capture acoustic information to facilitate recognition. Also, our cross-modal KD is specially designed for the Transformer structure, different from the previous work focused on CNN structure.

\section{Conclusion}
In this paper, we present \emph{Radio2Text}, a system that uses mmWave signals to perform streaming speech recognition with a vocabulary of over 13,000 words. We solve the challenges caused by low quality input mmWave signals and limitations of streaming ASR setting through the introduction of \emph{GI} and \emph{cross-modal KD}. Experimental results show that our Radio2Text can achieve the CER of 5.7\% and the WER of 9.4\% for the recognition of a vocabulary consisting of over 13,000 words, and it outperforms compared mmWave-based non-streaming ASR methods in quiet, noisy and soundproof scenarios. Moreover, in the quiet scenario, Radio2Text is on par with the microphone-based streaming ASR method, while in noisy and soundproof scenarios where the microphone-based method fails, \emph{Radio2Text} remains effective. We believe our \emph{Radio2Text} provides more possibilities for the application of mmWave-based speech recognition in real-world scenarios.

\begin{acks}
We thank the reviewers and editors for their constructive suggestions. We also thank Shiduo Zhang and Qiao Sun for their help with data collection. The authors are grateful to Rui Zhou, Yun-Hin Chan, Handi Chen, and Shizhen Zhao for discussion. 
\end{acks}

\bibliographystyle{ACM-Reference-Format}
\bibliography{reference}


\begin{thebibliography}{79}


\ifx \showCODEN    \undefined \def \showCODEN     #1{\unskip}     \fi
\ifx \showDOI      \undefined \def \showDOI       #1{#1}\fi
\ifx \showISBNx    \undefined \def \showISBNx     #1{\unskip}     \fi
\ifx \showISBNxiii \undefined \def \showISBNxiii  #1{\unskip}     \fi
\ifx \showISSN     \undefined \def \showISSN      #1{\unskip}     \fi
\ifx \showLCCN     \undefined \def \showLCCN      #1{\unskip}     \fi
\ifx \shownote     \undefined \def \shownote      #1{#1}          \fi
\ifx \showarticletitle \undefined \def \showarticletitle #1{#1}   \fi
\ifx \showURL      \undefined \def \showURL       {\relax}        \fi
\providecommand\bibfield[2]{#2}
\providecommand\bibinfo[2]{#2}
\providecommand\natexlab[1]{#1}
\providecommand\showeprint[2][]{arXiv:#2}

\bibitem[Afouras et~al\mbox{.}(2020)]%
        {asrallyouneed}
\bibfield{author}{\bibinfo{person}{Triantafyllos Afouras},
  \bibinfo{person}{Joon~Son Chung}, {and} \bibinfo{person}{Andrew Zisserman}.}
  \bibinfo{year}{2020}\natexlab{}.
\newblock \showarticletitle{ASR is All You Need: Cross-Modal Distillation for
  Lip Reading}. In \bibinfo{booktitle}{\emph{2020 IEEE International Conference
  on Acoustics, Speech and Signal Processing (ICASSP)}}.
  \bibinfo{pages}{2143--2147}.
\newblock
\urldef\tempurl%
\url{https://doi.org/10.1109/ICASSP40776.2020.9054253}
\showDOI{\tempurl}


\bibitem[Arivazhagan et~al\mbox{.}(2019)]%
        {streamingaed}
\bibfield{author}{\bibinfo{person}{Naveen Arivazhagan}, \bibinfo{person}{Colin
  Cherry}, \bibinfo{person}{Wolfgang Macherey}, \bibinfo{person}{Chung-Cheng
  Chiu}, \bibinfo{person}{Semih Yavuz}, \bibinfo{person}{Ruoming Pang},
  \bibinfo{person}{Wei Li}, {and} \bibinfo{person}{Colin Raffel}.}
  \bibinfo{year}{2019}\natexlab{}.
\newblock \showarticletitle{Monotonic Infinite Lookback Attention for
  Simultaneous Machine Translation}. In \bibinfo{booktitle}{\emph{Proceedings
  of the 57th Annual Meeting of the Association for Computational Linguistics
  (ACL)}}. \bibinfo{address}{Florence, Italy}, \bibinfo{pages}{1313--1323}.
\newblock
\urldef\tempurl%
\url{https://doi.org/10.18653/v1/P19-1126}
\showDOI{\tempurl}


\bibitem[Aytar et~al\mbox{.}(2016)]%
        {soundnet}
\bibfield{author}{\bibinfo{person}{Yusuf Aytar}, \bibinfo{person}{Carl
  Vondrick}, {and} \bibinfo{person}{Antonio Torralba}.}
  \bibinfo{year}{2016}\natexlab{}.
\newblock \showarticletitle{SoundNet: Learning Sound Representations from
  Unlabeled Video}. In \bibinfo{booktitle}{\emph{Advances in Neural Information
  Processing Systems (NeurIPS)}}, \bibfield{editor}{\bibinfo{person}{D.~Lee},
  \bibinfo{person}{M.~Sugiyama}, \bibinfo{person}{U.~Luxburg},
  \bibinfo{person}{I.~Guyon}, {and} \bibinfo{person}{R.~Garnett}} (Eds.),
  Vol.~\bibinfo{volume}{29}.
\newblock


\bibitem[Baevski et~al\mbox{.}(2020)]%
        {wav2vec2}
\bibfield{author}{\bibinfo{person}{Alexei Baevski}, \bibinfo{person}{Yuhao
  Zhou}, \bibinfo{person}{Abdelrahman Mohamed}, {and} \bibinfo{person}{Michael
  Auli}.} \bibinfo{year}{2020}\natexlab{}.
\newblock \showarticletitle{wav2vec 2.0: A framework for self-supervised
  learning of speech representations}.
\newblock \bibinfo{journal}{\emph{Advances in neural information processing
  systems (NeurIPS)}}  \bibinfo{volume}{33} (\bibinfo{year}{2020}),
  \bibinfo{pages}{12449--12460}.
\newblock


\bibitem[Basak and Gowda(2022)]%
        {mmspy}
\bibfield{author}{\bibinfo{person}{Suryoday Basak} {and}
  \bibinfo{person}{Mahanth Gowda}.} \bibinfo{year}{2022}\natexlab{}.
\newblock \showarticletitle{mmspy: Spying phone calls using mmwave radars}. In
  \bibinfo{booktitle}{\emph{2022 IEEE Symposium on Security and Privacy (SP)}}.
  IEEE, \bibinfo{pages}{1211--1228}.
\newblock


\bibitem[Becker et~al\mbox{.}(2018)]%
        {audiomnist}
\bibfield{author}{\bibinfo{person}{S{\"o}ren Becker}, \bibinfo{person}{Marcel
  Ackermann}, \bibinfo{person}{Sebastian Lapuschkin},
  \bibinfo{person}{Klaus-Robert M{\"u}ller}, {and} \bibinfo{person}{Wojciech
  Samek}.} \bibinfo{year}{2018}\natexlab{}.
\newblock \showarticletitle{Interpreting and explaining deep neural networks
  for classification of audio signals}.
\newblock \bibinfo{journal}{\emph{arXiv preprint arXiv:1807.03418}}
  (\bibinfo{year}{2018}).
\newblock


\bibitem[Bengio et~al\mbox{.}(2013)]%
        {Representationlearning}
\bibfield{author}{\bibinfo{person}{Yoshua Bengio}, \bibinfo{person}{Aaron
  Courville}, {and} \bibinfo{person}{Pascal Vincent}.}
  \bibinfo{year}{2013}\natexlab{}.
\newblock \showarticletitle{Representation learning: A review and new
  perspectives}.
\newblock \bibinfo{journal}{\emph{IEEE transactions on pattern analysis and
  machine intelligence}} \bibinfo{volume}{35}, \bibinfo{number}{8}
  (\bibinfo{year}{2013}), \bibinfo{pages}{1798--1828}.
\newblock


\bibitem[Benzeghiba et~al\mbox{.}(2007)]%
        {asrreview}
\bibfield{author}{\bibinfo{person}{M. Benzeghiba}, \bibinfo{person}{R. {De
  Mori}}, \bibinfo{person}{O. Deroo}, \bibinfo{person}{S. Dupont},
  \bibinfo{person}{T. Erbes}, \bibinfo{person}{D. Jouvet}, \bibinfo{person}{L.
  Fissore}, \bibinfo{person}{P. Laface}, \bibinfo{person}{A. Mertins},
  \bibinfo{person}{C. Ris}, \bibinfo{person}{R. Rose}, \bibinfo{person}{V.
  Tyagi}, {and} \bibinfo{person}{C. Wellekens}.}
  \bibinfo{year}{2007}\natexlab{}.
\newblock \showarticletitle{Automatic speech recognition and speech
  variability: A review}.
\newblock \bibinfo{journal}{\emph{Speech Communication}} \bibinfo{volume}{49},
  \bibinfo{number}{10} (\bibinfo{year}{2007}), \bibinfo{pages}{763--786}.
\newblock
\showISSN{0167-6393}
\urldef\tempurl%
\url{https://doi.org/10.1016/j.specom.2007.02.006}
\showDOI{\tempurl}


\bibitem[Cao and Liu(1996)]%
        {blindsourceseparation}
\bibfield{author}{\bibinfo{person}{Xi-Ren Cao} {and} \bibinfo{person}{Ruey-wen
  Liu}.} \bibinfo{year}{1996}\natexlab{}.
\newblock \showarticletitle{General approach to blind source separation}.
\newblock \bibinfo{journal}{\emph{IEEE Transactions on signal Processing}}
  \bibinfo{volume}{44}, \bibinfo{number}{3} (\bibinfo{year}{1996}),
  \bibinfo{pages}{562--571}.
\newblock


\bibitem[Chen et~al\mbox{.}(2017)]%
        {objectdetection1}
\bibfield{author}{\bibinfo{person}{Guobin Chen}, \bibinfo{person}{Wongun Choi},
  \bibinfo{person}{Xiang Yu}, \bibinfo{person}{Tony Han}, {and}
  \bibinfo{person}{Manmohan Chandraker}.} \bibinfo{year}{2017}\natexlab{}.
\newblock \showarticletitle{Learning efficient object detection models with
  knowledge distillation}.
\newblock \bibinfo{journal}{\emph{Advances in neural information processing
  systems (NeurIPS)}}  \bibinfo{volume}{30} (\bibinfo{year}{2017}).
\newblock


\bibitem[Chen et~al\mbox{.}(2021)]%
        {chunksetting}
\bibfield{author}{\bibinfo{person}{Xie Chen}, \bibinfo{person}{Yu Wu},
  \bibinfo{person}{Zhenghao Wang}, \bibinfo{person}{Shujie Liu}, {and}
  \bibinfo{person}{Jinyu Li}.} \bibinfo{year}{2021}\natexlab{}.
\newblock \showarticletitle{Developing Real-Time Streaming Transformer
  Transducer for Speech Recognition on Large-Scale Dataset}. In
  \bibinfo{booktitle}{\emph{ICASSP 2021 - 2021 IEEE International Conference on
  Acoustics, Speech and Signal Processing (ICASSP)}}.
  \bibinfo{pages}{5904--5908}.
\newblock
\urldef\tempurl%
\url{https://doi.org/10.1109/ICASSP39728.2021.9413535}
\showDOI{\tempurl}


\bibitem[Cherry(1953)]%
        {cocktail}
\bibfield{author}{\bibinfo{person}{E~Colin Cherry}.}
  \bibinfo{year}{1953}\natexlab{}.
\newblock \showarticletitle{Some experiments on the recognition of speech, with
  one and with two ears}.
\newblock \bibinfo{journal}{\emph{The Journal of the acoustical society of
  America}} \bibinfo{volume}{25}, \bibinfo{number}{5} (\bibinfo{year}{1953}),
  \bibinfo{pages}{975--979}.
\newblock


\bibitem[Clark et~al\mbox{.}(2019)]%
        {attentionbert}
\bibfield{author}{\bibinfo{person}{Kevin Clark}, \bibinfo{person}{Urvashi
  Khandelwal}, \bibinfo{person}{Omer Levy}, {and}
  \bibinfo{person}{Christopher~D. Manning}.} \bibinfo{year}{2019}\natexlab{}.
\newblock \showarticletitle{What Does {BERT} Look at? An Analysis of {BERT}{'}s
  Attention}. In \bibinfo{booktitle}{\emph{Proceedings of the 2019 ACL Workshop
  BlackboxNLP: Analyzing and Interpreting Neural Networks for NLP}}.
  \bibinfo{address}{Florence, Italy}, \bibinfo{pages}{276--286}.
\newblock
\urldef\tempurl%
\url{https://doi.org/10.18653/v1/W19-4828}
\showDOI{\tempurl}


\bibitem[Davis et~al\mbox{.}(2014)]%
        {visualmicrophone}
\bibfield{author}{\bibinfo{person}{Abe Davis}, \bibinfo{person}{Michael
  Rubinstein}, \bibinfo{person}{Neal Wadhwa}, \bibinfo{person}{Gautham~J.
  Mysore}, \bibinfo{person}{Fr\'{e}do Durand}, {and}
  \bibinfo{person}{William~T. Freeman}.} \bibinfo{year}{2014}\natexlab{}.
\newblock \showarticletitle{The Visual Microphone: Passive Recovery of Sound
  from Video}.
\newblock \bibinfo{journal}{\emph{ACM Trans. Graph.}} \bibinfo{volume}{33},
  \bibinfo{number}{4}, Article \bibinfo{articleno}{79} (\bibinfo{date}{jul}
  \bibinfo{year}{2014}), \bibinfo{numpages}{10}~pages.
\newblock
\showISSN{0730-0301}
\urldef\tempurl%
\url{https://doi.org/10.1145/2601097.2601119}
\showDOI{\tempurl}


\bibitem[Demonte(2019)]%
        {harvard}
\bibfield{author}{\bibinfo{person}{Philippa Demonte}.}
  \bibinfo{year}{2019}\natexlab{}.
\newblock \showarticletitle{HARVARD speech corpus--audio recording 2019}.
\newblock \bibinfo{journal}{\emph{University of Salford Collection}}
  (\bibinfo{year}{2019}).
\newblock


\bibitem[Fan et~al\mbox{.}(2020)]%
        {rfdiary}
\bibfield{author}{\bibinfo{person}{Lijie Fan}, \bibinfo{person}{Tianhong Li},
  \bibinfo{person}{Yuan Yuan}, {and} \bibinfo{person}{Dina Katabi}.}
  \bibinfo{year}{2020}\natexlab{}.
\newblock \showarticletitle{In-home daily-life captioning using radio signals}.
  In \bibinfo{booktitle}{\emph{Computer Vision--ECCV 2020: 16th European
  Conference}}. \bibinfo{pages}{105--123}.
\newblock


\bibitem[Fan et~al\mbox{.}(2023)]%
        {mmmic}
\bibfield{author}{\bibinfo{person}{Long Fan}, \bibinfo{person}{Lei Xie},
  \bibinfo{person}{Xinran Lu}, \bibinfo{person}{Yi Li}, \bibinfo{person}{Chuyu
  Wang}, {and} \bibinfo{person}{Sanglu Lu}.} \bibinfo{year}{2023}\natexlab{}.
\newblock \showarticletitle{mmmic: Multi-modal speech recognition based on
  mmwave radar}. In \bibinfo{booktitle}{\emph{The 42nd International IEEE
  Conference on Computer Communications (INFOCOM)}}.
\newblock


\bibitem[Feng et~al\mbox{.}(2023)]%
        {mmEavesdropper}
\bibfield{author}{\bibinfo{person}{Yiwen Feng}, \bibinfo{person}{Kai Zhang},
  \bibinfo{person}{Chuyu Wang}, \bibinfo{person}{Lei Xie},
  \bibinfo{person}{Jingyi Ning}, {and} \bibinfo{person}{Shijia Chen}.}
  \bibinfo{year}{2023}\natexlab{}.
\newblock \showarticletitle{mmEavesdropper: Signal Augmentation-based
  Directional Eavesdropping with mmWave Radar}. In
  \bibinfo{booktitle}{\emph{IEEE INFOCOM 2023 - IEEE Conference on Computer
  Communications}}.
\newblock


\bibitem[Gao et~al\mbox{.}(2022)]%
        {inertiear}
\bibfield{author}{\bibinfo{person}{Ming Gao}, \bibinfo{person}{Yajie Liu},
  \bibinfo{person}{Yike Chen}, \bibinfo{person}{Yimin Li},
  \bibinfo{person}{Zhongjie Ba}, \bibinfo{person}{Xian Xu}, {and}
  \bibinfo{person}{Jinsong Han}.} \bibinfo{year}{2022}\natexlab{}.
\newblock \showarticletitle{InertiEAR: Automatic and Device-independent
  IMU-based Eavesdropping on Smartphones}. In \bibinfo{booktitle}{\emph{IEEE
  INFOCOM 2022-IEEE Conference on Computer Communications}}. IEEE,
  \bibinfo{pages}{1129--1138}.
\newblock


\bibitem[Graves(2012)]%
        {rnnt}
\bibfield{author}{\bibinfo{person}{Alex Graves}.}
  \bibinfo{year}{2012}\natexlab{}.
\newblock \showarticletitle{Sequence transduction with recurrent neural
  networks}.
\newblock \bibinfo{journal}{\emph{arXiv preprint arXiv:1211.3711}}
  (\bibinfo{year}{2012}).
\newblock


\bibitem[Graves et~al\mbox{.}(2006)]%
        {ctc}
\bibfield{author}{\bibinfo{person}{Alex Graves}, \bibinfo{person}{Santiago
  Fern{\'a}ndez}, \bibinfo{person}{Faustino Gomez}, {and}
  \bibinfo{person}{J{\"u}rgen Schmidhuber}.} \bibinfo{year}{2006}\natexlab{}.
\newblock \showarticletitle{Connectionist temporal classification: labelling
  unsegmented sequence data with recurrent neural networks}. In
  \bibinfo{booktitle}{\emph{Proceedings of the 23rd international conference on
  Machine learning}}. \bibinfo{pages}{369--376}.
\newblock


\bibitem[Gulati et~al\mbox{.}(2020)]%
        {conformer}
\bibfield{author}{\bibinfo{person}{Anmol Gulati}, \bibinfo{person}{James Qin},
  \bibinfo{person}{Chung-Cheng Chiu}, \bibinfo{person}{Niki Parmar},
  \bibinfo{person}{Yu Zhang}, \bibinfo{person}{Jiahui Yu}, \bibinfo{person}{Wei
  Han}, \bibinfo{person}{Shibo Wang}, \bibinfo{person}{Zhengdong Zhang},
  \bibinfo{person}{Yonghui Wu}, {and} \bibinfo{person}{Ruoming Pang}.}
  \bibinfo{year}{2020}\natexlab{}.
\newblock \showarticletitle{{Conformer: Convolution-augmented Transformer for
  Speech Recognition}}. In \bibinfo{booktitle}{\emph{Proc. Interspeech 2020}}.
  \bibinfo{pages}{5036--5040}.
\newblock
\urldef\tempurl%
\url{https://doi.org/10.21437/Interspeech.2020-3015}
\showDOI{\tempurl}


\bibitem[Gupta et~al\mbox{.}(2016)]%
        {depthkd}
\bibfield{author}{\bibinfo{person}{Saurabh Gupta}, \bibinfo{person}{Judy
  Hoffman}, {and} \bibinfo{person}{Jitendra Malik}.}
  \bibinfo{year}{2016}\natexlab{}.
\newblock \showarticletitle{Cross modal distillation for supervision transfer}.
  In \bibinfo{booktitle}{\emph{Proceedings of the IEEE conference on computer
  vision and pattern recognition}}. \bibinfo{pages}{2827--2836}.
\newblock


\bibitem[Hannun et~al\mbox{.}(2014)]%
        {prefixbeamsearch}
\bibfield{author}{\bibinfo{person}{Awni~Y Hannun}, \bibinfo{person}{Andrew~L
  Maas}, \bibinfo{person}{Daniel Jurafsky}, {and} \bibinfo{person}{Andrew~Y
  Ng}.} \bibinfo{year}{2014}\natexlab{}.
\newblock \showarticletitle{First-pass large vocabulary continuous speech
  recognition using bi-directional recurrent dnns}.
\newblock \bibinfo{journal}{\emph{arXiv preprint arXiv:1408.2873}}
  (\bibinfo{year}{2014}).
\newblock


\bibitem[Hinton et~al\mbox{.}(2015)]%
        {kd}
\bibfield{author}{\bibinfo{person}{Geoffrey Hinton}, \bibinfo{person}{Oriol
  Vinyals}, {and} \bibinfo{person}{Jeffrey Dean}.}
  \bibinfo{year}{2015}\natexlab{}.
\newblock \showarticletitle{Distilling the Knowledge in a Neural Network}. In
  \bibinfo{booktitle}{\emph{NIPS Deep Learning and Representation Learning
  Workshop}}.
\newblock
\urldef\tempurl%
\url{http://arxiv.org/abs/1503.02531}
\showURL{%
\tempurl}


\bibitem[Hsu et~al\mbox{.}(2021)]%
        {hubert}
\bibfield{author}{\bibinfo{person}{Wei-Ning Hsu}, \bibinfo{person}{Benjamin
  Bolte}, \bibinfo{person}{Yao-Hung~Hubert Tsai}, \bibinfo{person}{Kushal
  Lakhotia}, \bibinfo{person}{Ruslan Salakhutdinov}, {and}
  \bibinfo{person}{Abdelrahman Mohamed}.} \bibinfo{year}{2021}\natexlab{}.
\newblock \showarticletitle{Hubert: Self-supervised speech representation
  learning by masked prediction of hidden units}.
\newblock \bibinfo{journal}{\emph{IEEE/ACM Transactions on Audio, Speech, and
  Language Processing}}  \bibinfo{volume}{29} (\bibinfo{year}{2021}),
  \bibinfo{pages}{3451--3460}.
\newblock


\bibitem[Hu et~al\mbox{.}(2020)]%
        {rnnttraining}
\bibfield{author}{\bibinfo{person}{Hu Hu}, \bibinfo{person}{Rui Zhao},
  \bibinfo{person}{Jinyu Li}, \bibinfo{person}{Liang Lu}, {and}
  \bibinfo{person}{Yifan Gong}.} \bibinfo{year}{2020}\natexlab{}.
\newblock \showarticletitle{Exploring Pre-Training with Alignments for RNN
  Transducer Based End-to-End Speech Recognition}. In
  \bibinfo{booktitle}{\emph{IEEE International Conference on Acoustics, Speech
  and Signal Processing (ICASSP)}}. \bibinfo{pages}{7079--7083}.
\newblock
\urldef\tempurl%
\url{https://doi.org/10.1109/ICASSP40776.2020.9054663}
\showDOI{\tempurl}


\bibitem[Hu et~al\mbox{.}(2022a)]%
        {mmecho}
\bibfield{author}{\bibinfo{person}{Pengfei Hu}, \bibinfo{person}{Wenhao Li},
  \bibinfo{person}{Riccardo Spolaor}, {and} \bibinfo{person}{Xiuzhen Cheng}.}
  \bibinfo{year}{2022}\natexlab{a}.
\newblock \showarticletitle{mmEcho: A mmWave-based Acoustic Eavesdropping
  Method}. In \bibinfo{booktitle}{\emph{2023 IEEE Symposium on Security and
  Privacy (SP)}}. IEEE Computer Society, \bibinfo{pages}{836--852}.
\newblock


\bibitem[Hu et~al\mbox{.}(2022b)]%
        {milliear}
\bibfield{author}{\bibinfo{person}{Pengfei Hu}, \bibinfo{person}{Yifan Ma},
  \bibinfo{person}{Panneer~Selvam Santhalingam}, \bibinfo{person}{Parth~H
  Pathak}, {and} \bibinfo{person}{Xiuzhen Cheng}.}
  \bibinfo{year}{2022}\natexlab{b}.
\newblock \showarticletitle{Milliear: Millimeter-wave acoustic eavesdropping
  with unconstrained vocabulary}. In \bibinfo{booktitle}{\emph{IEEE INFOCOM
  2022-IEEE Conference on Computer Communications}}. IEEE,
  \bibinfo{pages}{11--20}.
\newblock


\bibitem[Hu et~al\mbox{.}(2022c)]%
        {accear}
\bibfield{author}{\bibinfo{person}{Pengfei Hu}, \bibinfo{person}{Hui Zhuang},
  \bibinfo{person}{Panneer~Selvam Santhalingam}, \bibinfo{person}{Riccardo
  Spolaor}, \bibinfo{person}{Parth Pathak}, \bibinfo{person}{Guoming Zhang},
  {and} \bibinfo{person}{Xiuzhen Cheng}.} \bibinfo{year}{2022}\natexlab{c}.
\newblock \showarticletitle{Accear: Accelerometer acoustic eavesdropping with
  unconstrained vocabulary}. In \bibinfo{booktitle}{\emph{2022 IEEE Symposium
  on Security and Privacy (SP)}}. IEEE, \bibinfo{pages}{1757--1773}.
\newblock


\bibitem[Incorporated(2020a)]%
        {awr1642}
\bibfield{author}{\bibinfo{person}{Texas~Instruments Incorporated}.}
  \bibinfo{year}{2020}\natexlab{a}.
\newblock \bibinfo{booktitle}{\emph{AWR1642: Single-chip 76-GHz to 81-GHz
  automotive radar sensor integrating DSP and MCU}}.
\newblock
\urldef\tempurl%
\url{https://www.ti.com/product/AWR1642}
\showURL{%
\tempurl}


\bibitem[Incorporated(2020b)]%
        {dca1000evm}
\bibfield{author}{\bibinfo{person}{Texas~Instruments Incorporated}.}
  \bibinfo{year}{2020}\natexlab{b}.
\newblock \bibinfo{booktitle}{\emph{DCA1000EVM: Real-time data-capture adapter
  for radar sensing evaluation module}}.
\newblock
\urldef\tempurl%
\url{https://www.ti.com/tool/DCA1000EVM}
\showURL{%
\tempurl}


\bibitem[Islam* et~al\mbox{.}(2020)]%
        {position}
\bibfield{author}{\bibinfo{person}{Md~Amirul Islam*}, \bibinfo{person}{Sen
  Jia*}, {and} \bibinfo{person}{Neil D.~B. Bruce}.}
  \bibinfo{year}{2020}\natexlab{}.
\newblock \showarticletitle{How much Position Information Do Convolutional
  Neural Networks Encode?}. In \bibinfo{booktitle}{\emph{International
  Conference on Learning Representations (ICLR)}}.
\newblock


\bibitem[Ito and Johnson(2017)]%
        {ljspeech}
\bibfield{author}{\bibinfo{person}{Keith Ito} {and} \bibinfo{person}{Linda
  Johnson}.} \bibinfo{year}{2017}\natexlab{}.
\newblock \bibinfo{title}{The LJ Speech Dataset}.
\newblock
  \bibinfo{howpublished}{\url{https://keithito.com/LJ-Speech-Dataset/}}.
\newblock


\bibitem[Jardak et~al\mbox{.}(2019)]%
        {jardak2019compact}
\bibfield{author}{\bibinfo{person}{Seifallah Jardak},
  \bibinfo{person}{Mohamed-Slim Alouini}, \bibinfo{person}{Tero Kiuru},
  \bibinfo{person}{Mikko Metso}, {and} \bibinfo{person}{Sajid Ahmed}.}
  \bibinfo{year}{2019}\natexlab{}.
\newblock \showarticletitle{Compact mmWave FMCW radar: Implementation and
  performance analysis}.
\newblock \bibinfo{journal}{\emph{IEEE Aerospace and Electronic Systems
  Magazine}} \bibinfo{volume}{34}, \bibinfo{number}{2} (\bibinfo{year}{2019}),
  \bibinfo{pages}{36--44}.
\newblock


\bibitem[Jiang et~al\mbox{.}(2020)]%
        {mmvib}
\bibfield{author}{\bibinfo{person}{Chengkun Jiang}, \bibinfo{person}{Junchen
  Guo}, \bibinfo{person}{Yuan He}, \bibinfo{person}{Meng Jin},
  \bibinfo{person}{Shuai Li}, {and} \bibinfo{person}{Yunhao Liu}.}
  \bibinfo{year}{2020}\natexlab{}.
\newblock \showarticletitle{mmVib: micrometer-level vibration measurement with
  mmwave radar}. In \bibinfo{booktitle}{\emph{Proceedings of the 26th Annual
  International Conference on Mobile Computing and Networking}}.
  \bibinfo{pages}{1--13}.
\newblock


\bibitem[Khamis et~al\mbox{.}(2020)]%
        {RFWash}
\bibfield{author}{\bibinfo{person}{Abdelwahed Khamis},
  \bibinfo{person}{Branislav Kusy}, \bibinfo{person}{Chun~Tung Chou},
  \bibinfo{person}{Mary-Louise McLaws}, {and} \bibinfo{person}{Wen Hu}.}
  \bibinfo{year}{2020}\natexlab{}.
\newblock \showarticletitle{RFWash: a weakly supervised tracking of hand
  hygiene technique}. In \bibinfo{booktitle}{\emph{Proceedings of the 18th
  Conference on Embedded Networked Sensor Systems}}. \bibinfo{pages}{572--584}.
\newblock


\bibitem[Kim et~al\mbox{.}(2017)]%
        {jointctcattention}
\bibfield{author}{\bibinfo{person}{Suyoun Kim}, \bibinfo{person}{Takaaki Hori},
  {and} \bibinfo{person}{Shinji Watanabe}.} \bibinfo{year}{2017}\natexlab{}.
\newblock \showarticletitle{Joint CTC-attention based end-to-end speech
  recognition using multi-task learning}. In \bibinfo{booktitle}{\emph{2017
  IEEE international conference on acoustics, speech and signal processing
  (ICASSP)}}. IEEE, \bibinfo{pages}{4835--4839}.
\newblock


\bibitem[Kornblith et~al\mbox{.}(2019)]%
        {layersimilarity}
\bibfield{author}{\bibinfo{person}{Simon Kornblith}, \bibinfo{person}{Mohammad
  Norouzi}, \bibinfo{person}{Honglak Lee}, {and} \bibinfo{person}{Geoffrey
  Hinton}.} \bibinfo{year}{2019}\natexlab{}.
\newblock \showarticletitle{Similarity of Neural Network Representations
  Revisited}. In \bibinfo{booktitle}{\emph{Proceedings of the 36th
  International Conference on Machine Learning (ICML)}},
  Vol.~\bibinfo{volume}{97}. \bibinfo{publisher}{PMLR},
  \bibinfo{pages}{3519--3529}.
\newblock


\bibitem[Kudo and Richardson(2018)]%
        {sentencepiece}
\bibfield{author}{\bibinfo{person}{Taku Kudo} {and} \bibinfo{person}{John
  Richardson}.} \bibinfo{year}{2018}\natexlab{}.
\newblock \showarticletitle{{S}entence{P}iece: A simple and language
  independent subword tokenizer and detokenizer for Neural Text Processing}. In
  \bibinfo{booktitle}{\emph{Proceedings of the 2018 Conference on Empirical
  Methods in Natural Language Processing: System Demonstrations (EMNLP)}}.
  \bibinfo{publisher}{Association for Computational Linguistics},
  \bibinfo{address}{Brussels, Belgium}, \bibinfo{pages}{66--71}.
\newblock


\bibitem[Kwong et~al\mbox{.}(2019)]%
        {hard}
\bibfield{author}{\bibinfo{person}{Andrew Kwong}, \bibinfo{person}{Wenyuan Xu},
  {and} \bibinfo{person}{Kevin Fu}.} \bibinfo{year}{2019}\natexlab{}.
\newblock \showarticletitle{Hard drive of hearing: Disks that eavesdrop with a
  synthesized microphone}. In \bibinfo{booktitle}{\emph{2019 IEEE symposium on
  security and privacy (SP)}}. IEEE, \bibinfo{pages}{905--919}.
\newblock


\bibitem[Liu et~al\mbox{.}(2019b)]%
        {surveywireless}
\bibfield{author}{\bibinfo{person}{Jian Liu}, \bibinfo{person}{Hongbo Liu},
  \bibinfo{person}{Yingying Chen}, \bibinfo{person}{Yan Wang}, {and}
  \bibinfo{person}{Chen Wang}.} \bibinfo{year}{2019}\natexlab{b}.
\newblock \showarticletitle{Wireless sensing for human activity: A survey}.
\newblock \bibinfo{journal}{\emph{IEEE Communications Surveys \& Tutorials}}
  \bibinfo{volume}{22}, \bibinfo{number}{3} (\bibinfo{year}{2019}),
  \bibinfo{pages}{1629--1645}.
\newblock


\bibitem[Liu et~al\mbox{.}(2019a)]%
        {semanticseg1}
\bibfield{author}{\bibinfo{person}{Yifan Liu}, \bibinfo{person}{Ke Chen},
  \bibinfo{person}{Chris Liu}, \bibinfo{person}{Zengchang Qin},
  \bibinfo{person}{Zhenbo Luo}, {and} \bibinfo{person}{Jingdong Wang}.}
  \bibinfo{year}{2019}\natexlab{a}.
\newblock \showarticletitle{Structured knowledge distillation for semantic
  segmentation}. In \bibinfo{booktitle}{\emph{Proceedings of the IEEE/CVF
  conference on computer vision and pattern recognition (CVPR)}}.
  \bibinfo{pages}{2604--2613}.
\newblock
\urldef\tempurl%
\url{https://doi.org/10.1109/CVPR.2019.00271}
\showDOI{\tempurl}


\bibitem[Michalevsky et~al\mbox{.}(2014)]%
        {gyrophone}
\bibfield{author}{\bibinfo{person}{Yan Michalevsky}, \bibinfo{person}{Dan
  Boneh}, {and} \bibinfo{person}{Gabi Nakibly}.}
  \bibinfo{year}{2014}\natexlab{}.
\newblock \showarticletitle{Gyrophone: Recognizing speech from gyroscope
  signals}. In \bibinfo{booktitle}{\emph{23rd USENIX Security Symposium (USENIX
  Security 14)}}. \bibinfo{pages}{1053--1067}.
\newblock


\bibitem[Moritz et~al\mbox{.}(2020)]%
        {tastreamingtransformer}
\bibfield{author}{\bibinfo{person}{Niko Moritz}, \bibinfo{person}{Takaaki
  Hori}, {and} \bibinfo{person}{Jonathan Le}.} \bibinfo{year}{2020}\natexlab{}.
\newblock \showarticletitle{Streaming Automatic Speech Recognition with the
  Transformer Model}. In \bibinfo{booktitle}{\emph{IEEE International
  Conference on Acoustics, Speech and Signal Processing (ICASSP)}}.
  \bibinfo{pages}{6074--6078}.
\newblock


\bibitem[Ozturk et~al\mbox{.}(2023)]%
        {radioses}
\bibfield{author}{\bibinfo{person}{Muhammed~Zahid Ozturk},
  \bibinfo{person}{Chenshu Wu}, \bibinfo{person}{Beibei Wang},
  \bibinfo{person}{Min Wu}, {and} \bibinfo{person}{KJ~Ray Liu}.}
  \bibinfo{year}{2023}\natexlab{}.
\newblock \showarticletitle{Radio SES: mmWave-Based Audioradio Speech
  Enhancement and Separation System}.
\newblock \bibinfo{journal}{\emph{IEEE/ACM Transactions on Audio, Speech, and
  Language Processing}}  \bibinfo{volume}{31} (\bibinfo{year}{2023}),
  \bibinfo{pages}{1333--1347}.
\newblock


\bibitem[Panayotov et~al\mbox{.}(2015)]%
        {librispeech}
\bibfield{author}{\bibinfo{person}{Vassil Panayotov}, \bibinfo{person}{Guoguo
  Chen}, \bibinfo{person}{Daniel Povey}, {and} \bibinfo{person}{Sanjeev
  Khudanpur}.} \bibinfo{year}{2015}\natexlab{}.
\newblock \showarticletitle{Librispeech: an asr corpus based on public domain
  audio books}. In \bibinfo{booktitle}{\emph{2015 IEEE international conference
  on acoustics, speech and signal processing (ICASSP)}}. IEEE,
  \bibinfo{pages}{5206--5210}.
\newblock


\bibitem[Pratap et~al\mbox{.}(2023)]%
        {1000language}
\bibfield{author}{\bibinfo{person}{Vineel Pratap}, \bibinfo{person}{Andros
  Tjandra}, \bibinfo{person}{Bowen Shi}, \bibinfo{person}{Paden Tomasello},
  \bibinfo{person}{Arun Babu}, \bibinfo{person}{Sayani Kundu},
  \bibinfo{person}{Ali Elkahky}, \bibinfo{person}{Zhaoheng Ni},
  \bibinfo{person}{Apoorv Vyas}, \bibinfo{person}{Maryam Fazel-Zarandi},
  {et~al\mbox{.}}} \bibinfo{year}{2023}\natexlab{}.
\newblock \showarticletitle{Scaling speech technology to 1,000+ languages}.
\newblock \bibinfo{journal}{\emph{arXiv preprint arXiv:2305.13516}}
  (\bibinfo{year}{2023}).
\newblock


\bibitem[Romero et~al\mbox{.}(2015)]%
        {FitNets}
\bibfield{author}{\bibinfo{person}{Adriana Romero}, \bibinfo{person}{Nicolas
  Ballas}, \bibinfo{person}{Samira~Ebrahimi Kahou}, \bibinfo{person}{Antoine
  Chassang}, \bibinfo{person}{Carlo Gatta}, {and} \bibinfo{person}{Yoshua
  Bengio}.} \bibinfo{year}{2015}\natexlab{}.
\newblock \showarticletitle{FitNets: Hints for Thin Deep Nets}. In
  \bibinfo{booktitle}{\emph{3rd International Conference on Learning
  Representations, {ICLR} 2015, San Diego, CA, USA, May 7-9, 2015, Conference
  Track Proceedings}}.
\newblock
\urldef\tempurl%
\url{http://arxiv.org/abs/1412.6550}
\showURL{%
\tempurl}


\bibitem[Roy and Roy~Choudhury(2016)]%
        {listening}
\bibfield{author}{\bibinfo{person}{Nirupam Roy} {and} \bibinfo{person}{Romit
  Roy~Choudhury}.} \bibinfo{year}{2016}\natexlab{}.
\newblock \showarticletitle{Listening through a vibration motor}. In
  \bibinfo{booktitle}{\emph{Proceedings of the 14th Annual International
  Conference on Mobile Systems, Applications, and Services}}.
  \bibinfo{pages}{57--69}.
\newblock


\bibitem[Sami et~al\mbox{.}(2020)]%
        {spying}
\bibfield{author}{\bibinfo{person}{Sriram Sami}, \bibinfo{person}{Yimin Dai},
  \bibinfo{person}{Sean Rui~Xiang Tan}, \bibinfo{person}{Nirupam Roy}, {and}
  \bibinfo{person}{Jun Han}.} \bibinfo{year}{2020}\natexlab{}.
\newblock \showarticletitle{Spying with your robot vacuum cleaner:
  eavesdropping via lidar sensors}. In \bibinfo{booktitle}{\emph{Proceedings of
  the 18th Conference on Embedded Networked Sensor Systems}}.
  \bibinfo{pages}{354--367}.
\newblock


\bibitem[Sanh et~al\mbox{.}(2019)]%
        {bertcompress2}
\bibfield{author}{\bibinfo{person}{Victor Sanh}, \bibinfo{person}{Lysandre
  Debut}, \bibinfo{person}{Julien Chaumond}, {and} \bibinfo{person}{Thomas
  Wolf}.} \bibinfo{year}{2019}\natexlab{}.
\newblock \showarticletitle{DistilBERT, a distilled version of BERT: smaller,
  faster, cheaper and lighter}.
\newblock \bibinfo{journal}{\emph{arXiv preprint arXiv:1910.01108}}
  (\bibinfo{year}{2019}).
\newblock


\bibitem[Stoica et~al\mbox{.}(2005)]%
        {crosscorrelation}
\bibfield{author}{\bibinfo{person}{Petre Stoica}, \bibinfo{person}{Randolph~L
  Moses}, {et~al\mbox{.}}} \bibinfo{year}{2005}\natexlab{}.
\newblock \bibinfo{booktitle}{\emph{Spectral analysis of signals}}.
  Vol.~\bibinfo{volume}{452}.
\newblock \bibinfo{publisher}{Pearson Prentice Hall Upper Saddle River, NJ}.
\newblock


\bibitem[Sun et~al\mbox{.}(2019)]%
        {bertcompress1}
\bibfield{author}{\bibinfo{person}{Siqi Sun}, \bibinfo{person}{Yu Cheng},
  \bibinfo{person}{Zhe Gan}, {and} \bibinfo{person}{Jingjing Liu}.}
  \bibinfo{year}{2019}\natexlab{}.
\newblock \showarticletitle{Patient Knowledge Distillation for {BERT} Model
  Compression}. In \bibinfo{booktitle}{\emph{Proceedings of the 2019 Conference
  on Empirical Methods in Natural Language Processing and the 9th International
  Joint Conference on Natural Language Processing (EMNLP-IJCNLP)}}.
  \bibinfo{address}{Hong Kong, China}, \bibinfo{pages}{4323--4332}.
\newblock
\urldef\tempurl%
\url{https://doi.org/10.18653/v1/D19-1441}
\showDOI{\tempurl}


\bibitem[Sun et~al\mbox{.}(2020)]%
        {fmcwautonomous}
\bibfield{author}{\bibinfo{person}{Shunqiao Sun}, \bibinfo{person}{Athina~P.
  Petropulu}, {and} \bibinfo{person}{H.~Vincent Poor}.}
  \bibinfo{year}{2020}\natexlab{}.
\newblock \showarticletitle{MIMO Radar for Advanced Driver-Assistance Systems
  and Autonomous Driving: Advantages and Challenges}.
\newblock \bibinfo{journal}{\emph{IEEE Signal Processing Magazine}}
  \bibinfo{volume}{37}, \bibinfo{number}{4} (\bibinfo{year}{2020}),
  \bibinfo{pages}{98--117}.
\newblock
\urldef\tempurl%
\url{https://doi.org/10.1109/MSP.2020.2978507}
\showDOI{\tempurl}


\bibitem[Tay et~al\mbox{.}(2022)]%
        {efficienttransformer}
\bibfield{author}{\bibinfo{person}{Yi Tay}, \bibinfo{person}{Mostafa Dehghani},
  \bibinfo{person}{Dara Bahri}, {and} \bibinfo{person}{Donald Metzler}.}
  \bibinfo{year}{2022}\natexlab{}.
\newblock \showarticletitle{Efficient Transformers: A Survey}.
\newblock \bibinfo{journal}{\emph{ACM Comput. Surv.}} \bibinfo{volume}{55},
  \bibinfo{number}{6}, Article \bibinfo{articleno}{109} (\bibinfo{date}{dec}
  \bibinfo{year}{2022}), \bibinfo{numpages}{28}~pages.
\newblock
\urldef\tempurl%
\url{https://doi.org/10.1145/3530811}
\showDOI{\tempurl}


\bibitem[Vaswani et~al\mbox{.}(2017)]%
        {transformer}
\bibfield{author}{\bibinfo{person}{Ashish Vaswani}, \bibinfo{person}{Noam
  Shazeer}, \bibinfo{person}{Niki Parmar}, \bibinfo{person}{Jakob Uszkoreit},
  \bibinfo{person}{Llion Jones}, \bibinfo{person}{Aidan~N Gomez},
  \bibinfo{person}{\L~ukasz Kaiser}, {and} \bibinfo{person}{Illia Polosukhin}.}
  \bibinfo{year}{2017}\natexlab{}.
\newblock \showarticletitle{Attention is All you Need}. In
  \bibinfo{booktitle}{\emph{Advances in Neural Information Processing Systems
  (NeurIPS)}}, Vol.~\bibinfo{volume}{30}.
\newblock


\bibitem[Wang et~al\mbox{.}(2022a)]%
        {wavesdropper}
\bibfield{author}{\bibinfo{person}{Chao Wang}, \bibinfo{person}{Feng Lin},
  \bibinfo{person}{Zhongjie Ba}, \bibinfo{person}{Fan Zhang},
  \bibinfo{person}{Wenyao Xu}, {and} \bibinfo{person}{Kui Ren}.}
  \bibinfo{year}{2022}\natexlab{a}.
\newblock \showarticletitle{Wavesdropper: Through-wall Word Detection of Human
  Speech via Commercial mmWave Devices}.
\newblock \bibinfo{journal}{\emph{Proceedings of the ACM on Interactive,
  Mobile, Wearable and Ubiquitous Technologies}} \bibinfo{volume}{6},
  \bibinfo{number}{2} (\bibinfo{year}{2022}), \bibinfo{pages}{1--26}.
\newblock


\bibitem[Wang et~al\mbox{.}(2022b)]%
        {mmphone}
\bibfield{author}{\bibinfo{person}{Chao Wang}, \bibinfo{person}{Feng Lin},
  \bibinfo{person}{Tiantian Liu}, \bibinfo{person}{Ziwei Liu},
  \bibinfo{person}{Yijie Shen}, \bibinfo{person}{Zhongjie Ba},
  \bibinfo{person}{Li Lu}, \bibinfo{person}{Wenyao Xu}, {and}
  \bibinfo{person}{Kui Ren}.} \bibinfo{year}{2022}\natexlab{b}.
\newblock \showarticletitle{mmphone: Acoustic eavesdropping on loudspeakers via
  mmwave-characterized piezoelectric effect}. In \bibinfo{booktitle}{\emph{IEEE
  INFOCOM 2022-IEEE Conference on Computer Communications}}. IEEE,
  \bibinfo{pages}{820--829}.
\newblock


\bibitem[Wang et~al\mbox{.}(2022c)]%
        {mmeve}
\bibfield{author}{\bibinfo{person}{Chao Wang}, \bibinfo{person}{Feng Lin},
  \bibinfo{person}{Tiantian Liu}, \bibinfo{person}{Kaidi Zheng},
  \bibinfo{person}{Zhibo Wang}, \bibinfo{person}{Zhengxiong Li},
  \bibinfo{person}{Ming-Chun Huang}, \bibinfo{person}{Wenyao Xu}, {and}
  \bibinfo{person}{Kui Ren}.} \bibinfo{year}{2022}\natexlab{c}.
\newblock \showarticletitle{mmEve: eavesdropping on smartphone's earpiece via
  COTS mmWave device}. In \bibinfo{booktitle}{\emph{Proceedings of the 28th
  Annual International Conference on Mobile Computing And Networking}}.
  \bibinfo{pages}{338--351}.
\newblock


\bibitem[Wang et~al\mbox{.}(2020)]%
        {scout}
\bibfield{author}{\bibinfo{person}{Chengyi Wang}, \bibinfo{person}{Yu Wu},
  \bibinfo{person}{Liang Lu}, \bibinfo{person}{Shujie Liu},
  \bibinfo{person}{Jinyu Li}, \bibinfo{person}{Guoli Ye}, {and}
  \bibinfo{person}{Ming Zhou}.} \bibinfo{year}{2020}\natexlab{}.
\newblock \showarticletitle{{Low Latency End-to-End Streaming Speech
  Recognition with a Scout Network}}. In \bibinfo{booktitle}{\emph{Proc.
  Interspeech 2020}}. \bibinfo{pages}{2112--2116}.
\newblock


\bibitem[Wang and Chen(2018)]%
        {dlseparation}
\bibfield{author}{\bibinfo{person}{DeLiang Wang} {and} \bibinfo{person}{Jitong
  Chen}.} \bibinfo{year}{2018}\natexlab{}.
\newblock \showarticletitle{Supervised speech separation based on deep
  learning: An overview}.
\newblock \bibinfo{journal}{\emph{IEEE/ACM Transactions on Audio, Speech, and
  Language Processing}} \bibinfo{volume}{26}, \bibinfo{number}{10}
  (\bibinfo{year}{2018}), \bibinfo{pages}{1702--1726}.
\newblock


\bibitem[Wang and Wang(2020)]%
        {interpolation}
\bibfield{author}{\bibinfo{person}{Heming Wang} {and} \bibinfo{person}{Deliang
  Wang}.} \bibinfo{year}{2020}\natexlab{}.
\newblock \showarticletitle{Time-frequency loss for CNN based speech
  super-resolution}. In \bibinfo{booktitle}{\emph{Proceedings of IEEE
  International Conference on Acoustics, Speech and Signal Processing
  (ICASSP)}}. IEEE, \bibinfo{pages}{861--865}.
\newblock


\bibitem[Warden(2018)]%
        {commands}
\bibfield{author}{\bibinfo{person}{Pete Warden}.}
  \bibinfo{year}{2018}\natexlab{}.
\newblock \showarticletitle{Speech commands: A dataset for limited-vocabulary
  speech recognition}.
\newblock \bibinfo{journal}{\emph{arXiv preprint arXiv:1804.03209}}
  (\bibinfo{year}{2018}).
\newblock


\bibitem[Watanabe et~al\mbox{.}(2018)]%
        {espnet}
\bibfield{author}{\bibinfo{person}{Shinji Watanabe}, \bibinfo{person}{Takaaki
  Hori}, \bibinfo{person}{Shigeki Karita}, \bibinfo{person}{Tomoki Hayashi},
  \bibinfo{person}{Jiro Nishitoba}, \bibinfo{person}{Yuya Unno},
  \bibinfo{person}{Nelson {Enrique Yalta Soplin}}, \bibinfo{person}{Jahn
  Heymann}, \bibinfo{person}{Matthew Wiesner}, \bibinfo{person}{Nanxin Chen},
  \bibinfo{person}{Adithya Renduchintala}, {and} \bibinfo{person}{Tsubasa
  Ochiai}.} \bibinfo{year}{2018}\natexlab{}.
\newblock \showarticletitle{{ESPnet: End-to-End Speech Processing Toolkit}}. In
  \bibinfo{booktitle}{\emph{Proc. Interspeech 2018}}.
  \bibinfo{pages}{2207--2211}.
\newblock


\bibitem[Wei et~al\mbox{.}(2015)]%
        {wei2015acoustic}
\bibfield{author}{\bibinfo{person}{Teng Wei}, \bibinfo{person}{Shu Wang},
  \bibinfo{person}{Anfu Zhou}, {and} \bibinfo{person}{Xinyu Zhang}.}
  \bibinfo{year}{2015}\natexlab{}.
\newblock \showarticletitle{Acoustic eavesdropping through wireless
  vibrometry}. In \bibinfo{booktitle}{\emph{Proceedings of the 21st Annual
  International Conference on Mobile Computing and Networking}}.
  \bibinfo{pages}{130--141}.
\newblock


\bibitem[Xu et~al\mbox{.}(2019)]%
        {waveear}
\bibfield{author}{\bibinfo{person}{Chenhan Xu}, \bibinfo{person}{Zhengxiong
  Li}, \bibinfo{person}{Hanbin Zhang}, \bibinfo{person}{Aditya~Singh Rathore},
  \bibinfo{person}{Huining Li}, \bibinfo{person}{Chen Song},
  \bibinfo{person}{Kun Wang}, {and} \bibinfo{person}{Wenyao Xu}.}
  \bibinfo{year}{2019}\natexlab{}.
\newblock \showarticletitle{WaveEar: Exploring a MmWave-Based Noise-Resistant
  Speech Sensing for Voice-User Interface}. In
  \bibinfo{booktitle}{\emph{Proceedings of the 17th Annual International
  Conference on Mobile Systems, Applications, and Services}}
  \emph{(\bibinfo{series}{MobiSys '19})}. \bibinfo{address}{New York, NY, USA},
  \bibinfo{pages}{14–26}.
\newblock


\bibitem[Xue et~al\mbox{.}(2021)]%
        {mmMesh}
\bibfield{author}{\bibinfo{person}{Hongfei Xue}, \bibinfo{person}{Yan Ju},
  \bibinfo{person}{Chenglin Miao}, \bibinfo{person}{Yijiang Wang},
  \bibinfo{person}{Shiyang Wang}, \bibinfo{person}{Aidong Zhang}, {and}
  \bibinfo{person}{Lu Su}.} \bibinfo{year}{2021}\natexlab{}.
\newblock \showarticletitle{mmMesh: Towards 3D Real-Time Dynamic Human Mesh
  Construction Using Millimeter-Wave}. In \bibinfo{booktitle}{\emph{Proceedings
  of the 19th Annual International Conference on Mobile Systems, Applications,
  and Services}} \emph{(\bibinfo{series}{MobiSys '21})}. \bibinfo{address}{New
  York, NY, USA}, \bibinfo{pages}{269–282}.
\newblock
\urldef\tempurl%
\url{https://doi.org/10.1145/3458864.3467679}
\showDOI{\tempurl}


\bibitem[Xue et~al\mbox{.}(2022)]%
        {multimodalitykdanalysis}
\bibfield{author}{\bibinfo{person}{Zihui Xue}, \bibinfo{person}{Zhengqi Gao},
  \bibinfo{person}{Sucheng Ren}, {and} \bibinfo{person}{Hang Zhao}.}
  \bibinfo{year}{2022}\natexlab{}.
\newblock \showarticletitle{The Modality Focusing Hypothesis: Towards
  Understanding Crossmodal Knowledge Distillation}. In
  \bibinfo{booktitle}{\emph{The Eleventh International Conference on Learning
  Representations (ICLR)}}.
\newblock


\bibitem[Yang et~al\mbox{.}(2022b)]%
        {semanticseg2}
\bibfield{author}{\bibinfo{person}{Chuanguang Yang}, \bibinfo{person}{Helong
  Zhou}, \bibinfo{person}{Zhulin An}, \bibinfo{person}{Xue Jiang},
  \bibinfo{person}{Yongjun Xu}, {and} \bibinfo{person}{Qian Zhang}.}
  \bibinfo{year}{2022}\natexlab{b}.
\newblock \showarticletitle{Cross-image relational knowledge distillation for
  semantic segmentation}. In \bibinfo{booktitle}{\emph{Proceedings of the
  IEEE/CVF Conference on Computer Vision and Pattern Recognition (CVPR)}}.
  \bibinfo{pages}{12319--12328}.
\newblock


\bibitem[Yang et~al\mbox{.}(2022a)]%
        {objectdetection2}
\bibfield{author}{\bibinfo{person}{Jihan Yang}, \bibinfo{person}{Shaoshuai
  Shi}, \bibinfo{person}{Runyu Ding}, \bibinfo{person}{Zhe Wang}, {and}
  \bibinfo{person}{Xiaojuan Qi}.} \bibinfo{year}{2022}\natexlab{a}.
\newblock \showarticletitle{Towards efficient 3d object detection with
  knowledge distillation}.
\newblock \bibinfo{journal}{\emph{Advances in Neural Information Processing
  Systems (NeurIPS)}}  \bibinfo{volume}{35} (\bibinfo{year}{2022}),
  \bibinfo{pages}{21300--21313}.
\newblock


\bibitem[Zagoruyko and Komodakis(2017)]%
        {AttentionTransfer}
\bibfield{author}{\bibinfo{person}{Sergey Zagoruyko} {and}
  \bibinfo{person}{Nikos Komodakis}.} \bibinfo{year}{2017}\natexlab{}.
\newblock \showarticletitle{Paying More Attention to Attention: Improving the
  Performance of Convolutional Neural Networks via Attention Transfer}. In
  \bibinfo{booktitle}{\emph{5th International Conference on Learning
  Representations, {ICLR} 2017, Toulon, France, April 24-26, 2017, Conference
  Track Proceedings}}.
\newblock


\bibitem[Zhang et~al\mbox{.}(2023)]%
        {surveymmwave}
\bibfield{author}{\bibinfo{person}{Jia Zhang}, \bibinfo{person}{Rui Xi},
  \bibinfo{person}{Yuan He}, \bibinfo{person}{Yimiao Sun},
  \bibinfo{person}{Xiuzhen Guo}, \bibinfo{person}{Weiguo Wang},
  \bibinfo{person}{Xin Na}, \bibinfo{person}{Yunhao Liu},
  \bibinfo{person}{Zhenguo Shi}, {and} \bibinfo{person}{Tao Gu}.}
  \bibinfo{year}{2023}\natexlab{}.
\newblock \showarticletitle{A Survey of mmWave-Based Human Sensing: Technology,
  Platforms and Applications}.
\newblock \bibinfo{journal}{\emph{IEEE Communications Surveys \& Tutorials}}
  (\bibinfo{year}{2023}), \bibinfo{pages}{1--1}.
\newblock
\urldef\tempurl%
\url{https://doi.org/10.1109/COMST.2023.3298300}
\showDOI{\tempurl}


\bibitem[Zhang et~al\mbox{.}(2022b)]%
        {ambiear}
\bibfield{author}{\bibinfo{person}{Jia Zhang}, \bibinfo{person}{Yinian Zhou},
  \bibinfo{person}{Rui Xi}, \bibinfo{person}{Shuai Li},
  \bibinfo{person}{Junchen Guo}, {and} \bibinfo{person}{Yuan He}.}
  \bibinfo{year}{2022}\natexlab{b}.
\newblock \showarticletitle{AmbiEar: mmWave Based Voice Recognition in NLoS
  Scenarios}.
\newblock \bibinfo{journal}{\emph{Proceedings of the ACM on Interactive,
  Mobile, Wearable and Ubiquitous Technologies}} \bibinfo{volume}{6},
  \bibinfo{number}{3} (\bibinfo{year}{2022}), \bibinfo{pages}{1--25}.
\newblock


\bibitem[Zhang et~al\mbox{.}(2022a)]%
        {heartbeat}
\bibfield{author}{\bibinfo{person}{Shujie Zhang}, \bibinfo{person}{Tianyue
  Zheng}, \bibinfo{person}{Zhe Chen}, {and} \bibinfo{person}{Jun Luo}.}
  \bibinfo{year}{2022}\natexlab{a}.
\newblock \showarticletitle{Can We Obtain Fine-grained Heartbeat Waveform via
  Contact-free RF-sensing?}. In \bibinfo{booktitle}{\emph{IEEE INFOCOM
  2022-IEEE Conference on Computer Communications}}. IEEE,
  \bibinfo{pages}{1759--1768}.
\newblock


\bibitem[Zhao et~al\mbox{.}(2018a)]%
        {humanpose}
\bibfield{author}{\bibinfo{person}{Mingmin Zhao}, \bibinfo{person}{Tianhong
  Li}, \bibinfo{person}{Mohammad Abu~Alsheikh}, \bibinfo{person}{Yonglong
  Tian}, \bibinfo{person}{Hang Zhao}, \bibinfo{person}{Antonio Torralba}, {and}
  \bibinfo{person}{Dina Katabi}.} \bibinfo{year}{2018}\natexlab{a}.
\newblock \showarticletitle{Through-wall human pose estimation using radio
  signals}. In \bibinfo{booktitle}{\emph{Proceedings of the IEEE Conference on
  Computer Vision and Pattern Recognition (CVPR)}}.
  \bibinfo{pages}{7356--7365}.
\newblock


\bibitem[Zhao et~al\mbox{.}(2018b)]%
        {rf3d}
\bibfield{author}{\bibinfo{person}{Mingmin Zhao}, \bibinfo{person}{Yonglong
  Tian}, \bibinfo{person}{Hang Zhao}, \bibinfo{person}{Mohammad~Abu Alsheikh},
  \bibinfo{person}{Tianhong Li}, \bibinfo{person}{Rumen Hristov},
  \bibinfo{person}{Zachary Kabelac}, \bibinfo{person}{Dina Katabi}, {and}
  \bibinfo{person}{Antonio Torralba}.} \bibinfo{year}{2018}\natexlab{b}.
\newblock \showarticletitle{RF-based 3D skeletons}. In
  \bibinfo{booktitle}{\emph{Proceedings of the 2018 Conference of the ACM
  Special Interest Group on Data Communication (SIGCOMM)}}.
  \bibinfo{pages}{267--281}.
\newblock


\bibitem[Zhao et~al\mbox{.}(2020)]%
        {end2endradar}
\bibfield{author}{\bibinfo{person}{Running Zhao}, \bibinfo{person}{Xiaolin Ma},
  \bibinfo{person}{Xinhua Liu}, {and} \bibinfo{person}{Jian Liu}.}
  \bibinfo{year}{2020}\natexlab{}.
\newblock \showarticletitle{An end-to-end network for continuous human motion
  recognition via radar radios}.
\newblock \bibinfo{journal}{\emph{IEEE Sensors Journal}} \bibinfo{volume}{21},
  \bibinfo{number}{5} (\bibinfo{year}{2020}), \bibinfo{pages}{6487--6496}.
\newblock


\bibitem[Zhao et~al\mbox{.}(2022)]%
        {radio2speech}
\bibfield{author}{\bibinfo{person}{Running Zhao}, \bibinfo{person}{Jiangtao
  Yu}, \bibinfo{person}{Tingle Li}, \bibinfo{person}{Hang Zhao}, {and}
  \bibinfo{person}{Edith C.~H. Ngai}.} \bibinfo{year}{2022}\natexlab{}.
\newblock \showarticletitle{{Radio2Speech: High Quality Speech Recovery from
  Radio Frequency Signals}}. In \bibinfo{booktitle}{\emph{Proc. Interspeech
  2022}}. \bibinfo{pages}{4666--4670}.
\newblock


\end{thebibliography}



\end{document}